\documentclass{aa}
\usepackage{txfonts}
\usepackage{graphicx}
\usepackage{natbib}
\usepackage{amssymb}
\usepackage{lscape}
\bibpunct{(}{)}{;}{a}{}{,}

\def\chandra{{\it Chandra~}}
\def\chandrak{{\it Chandra}}
\def\swift{{\it Swift~}}

\def\xmm{{\it XMM-Newton~}}
\def\xmmk{{\it XMM-Newton}}

\def\m31{{M~31}}
\def\msun{{$M_{\sun}$}}

\def\pe{PFF2005~}
\def\pz{PHS2007~}
\def\pek{PFF2005}
\def\pzk{PHS2007}
\def\me{Paper~I~}
\def\mek{Paper~I}
\def\mz{Paper~II~}
\def\mzk{Paper~II}
\def\mb{Papers~I~\&~II~}
\def\mbk{Papers~I~\&~II}

\newcommand{\nh}{\hbox{$N_{\rm H}$}~}
\newcommand{\hcm}[1]{$\times 10^{#1}$ cm$^{-2}$}
\newcommand{\ohcm}[1]{$10^{#1}$ cm$^{-2}$}

\newcommand{\tpower}[1]{$\times 10^{#1}$}

\newcommand{\justttwo}{$t_{2}$~}

\newcommand{\ton}{$t_{\mbox{\small{on}}}$~}
\newcommand{\toff}{$t_{\mbox{\small{off}}}$~}
\newcommand{\vexp}{$v_{\mbox{\small{exp}}}$~}
\newcommand{\ttwo}{$t_{2,R}$~}

\newcommand{\tonk}{$t_{\mbox{\small{on}}}$}
\newcommand{\toffk}{$t_{\mbox{\small{off}}}$}
\newcommand{\vexpk}{$v_{\mbox{\small{exp}}}$}
\newcommand{\ttwok}{$t_{2,R}$}

\newcommand{\eton}{t_{\mbox{\small{on}}}}
\newcommand{\etoff}{t_{\mbox{\small{off}}}}
\newcommand{\evexp}{v_{\mbox{\small{exp}}}}
\newcommand{\ettwo}{t_{2,R}}

\newcommand{\emburn}{M_{\mbox{\small{burn,H}}}}
\newcommand{\mburnk}{$M_{\mbox{\small{burn,H}}}$}

\newcommand{\mej}{$M_{\mbox{\small{ej,H}}}$~}
\newcommand{\emej}{M_{\mbox{\small{ej,H}}}}
\newcommand{\mejk}{$M_{\mbox{\small{ej,H}}}$}

\begin{document}

\title{X-ray monitoring of classical novae in the central region of \m31\\ III. Autumn and winter 2009/10, 2010/11, and 2011/12\thanks{Partly based on observations with \xmmk, an ESA Science Mission with instruments and contributions directly funded by ESA Member States and NASA}}

\author{M.~Henze\inst{1,2}
	\and W.~Pietsch\inst{1}
	\and F.~Haberl\inst{1}
	\and M.~Della Valle\inst{3,4}
  \and G.~Sala\inst{5,6}
  \and D.~Hatzidimitriou\inst{7}
  \and F.~Hofmann\inst{1}
  \and M.~Hernanz\inst{8}
  \and D.H.~Hartmann\inst{9}
	\and J.~Greiner\inst{1}
}

\institute{Max-Planck-Institut f\"ur extraterrestrische Physik, Giessenbachstra\ss e,
	D-85748 Garching, Germany
	\and European Space Astronomy Centre, P.O. Box 78, 28691 Villanueva de la Ca\~{n}ada, Madrid, Spain\\
	email: \texttt{mhenze@sciops.esa.int}
  \and INAF-Napoli, Osservatorio Astronomico di Capodimonte, Salita Moiariello 16, I-80131 Napoli, Italy
  \and International Centre for Relativistic Astrophysics, Piazzale della Repubblica 2, I-65122 Pescara, Italy
  \and Departament de F\'isica i Enginyeria Nuclear, EUETIB, Universitat Polit\`ecnica de Catalunya, c/ Comte d'Urgell 187, 08036 Barcelona, Spain
  \and Institut d'Estudis Espacials de Catalunya, c/Gran Capit\`a 2-4, Ed. Nexus-201, 08034, Barcelona, Spain
  \and Department of Astrophysics, Astronomy and Mechanics, Faculty of Physics, University of Athens, Panepistimiopolis, GR15784 Zografos, Athens, Greece
  \and Institut de Ci\`encies de l'Espai (CSIC-IEEC), Campus UAB, Fac. Ci\`encies, E-08193 Bellaterra, Spain
  \and Department of Physics and Astronomy, Clemson University, Clemson, SC 29634-0978, USA
}

\date{Received 1 August 2013 / Accepted 28 November 2013}

\abstract
{Classical novae (CNe) represent the major class of supersoft X-ray sources (SSSs) in the central region of our neighbouring galaxy \m31.}
{We performed a dedicated monitoring of the \m31 central region, which aimed to detect SSS counterparts of CNe, with \xmm and \chandra between Nov and Mar of the years 2009/10, 2010/11, and 2011/12.}
{We systematically searched our data for X-ray counterparts of CNe and determined their X-ray light curves and also their spectral properties in the case of \xmm data. Additionally, we determined luminosity upper limits for all previously known X-ray emitting novae, which are not detected anymore, and for all CNe in our field of view with recent optical outbursts.}
{In total, we detected 24 novae in X-rays. Seven of these sources were known from previous observations, including the \m31 nova with the longest SSS phase, M31N~1996-08b, which was found to fade below our X-ray detection limit 13.8~yr after outburst. Of the new discoveries, several novae exhibit significant variability in their short-term X-ray light curves with one object showing a suspected period of about 1.3~h. We studied the SSS state of the most recent outburst of a recurrent nova, which had previously shown the shortest time ever observed between two outbursts ($\sim 5$~yr). The total number of \m31 novae with X-ray counterpart was increased to 79, and we subjected this extended catalogue to detailed statistical studies. Four previously indicated correlations between optical and X-ray parameters could be confirmed and improved. Furthermore, we found indications that the multi-dimensional parameter space of nova properties might be dominated by a single physical parameter, and we provide interpretations and suggest implications. We studied various outliers from the established correlations and discuss evidence of a different X-ray behaviour of novae in the \m31 bulge and disk.}
{Exploration of the multi-wavelength parameter space of optical and X-ray measurements is shown to be a powerful tool for examining properties of extragalactic nova populations. While there are hints that the different stellar populations of \m31 (bulge vs disk) produce dissimilar nova outbursts, there is also growing evidence that the overall behaviour of an average nova might be understood in surprisingly simple terms.}


\titlerunning{X-ray monitoring of classical novae in \m31 in 2009/10, 2010/11, and 2011/12}

\maketitle

%
%
\section{Introduction}
\label{sec:intro}
%
This is the third in a series of papers analysing data from X-ray monitoring campaigns for classical novae (CNe) in the central region of our neighbour galaxy \m31. In the first two papers, we presented the results of earlier campaigns from Jun 2006 to Mar 2007 \citep[][hereafter \mek]{2010A&A...523A..89H}, Nov 2007 until Feb 2008, and Nov 2008 until Feb 2009 \citep[both in][hereafter \mzk]{2011A&A...533A..52H}. This work presents the results of another three monitoring seasons with \xmm and \chandra during the autumn and winter of the years 2009/10, 2010/11 and 2011/12.

Classical nova events occur in binary systems of the cataclysmic variable (CV) type, which are those experiencing mass transfer from the main sequence or red giant secondary star onto the primary white dwarf (WD) component of the system \citep[for recent reviews see][]{2008clno.book.....B}. The nova outburst is triggered by a thermonuclear runaway in the accreted (hydrogen) matter. The optical nova, the phenomenological discovery of a ``new star'' where none was known before, is the product of the rapidly expanding hot envelope creating a massively enlarged pseudo-photosphere within hours to a few days. At its optical maximum, a nova can be from seven to 16 magnitudes brighter than in quiescence \citep[see][and references therein]{2010AJ....140...34S}.

After the optical maximum, the CNs photosphere recedes towards inner, hotter layers and the peak of emission shifts to shorter wavelengths. The speed of decline of the optical magnitude is one of the main observable parameters of the CN outburst. \citet{1964gano.book.....P} first established \citep[but see also][]{1936PA.....44...78G,1939PA.....47..538M} a detailed system of nova speed classes based on the time (in days) needed for the CN light curve to decay by two or three magnitudes below maximum magnitude ($t_2$ or $t_3$). The decline times have been found to be connected to the CNs peak magnitude \citep[the maximum magnitude to rate of decline, or MMRD, relationship; see][]{1995ApJ...452..704D} and to the expansion velocity (\vexpk) of the ejected envelope \citep{2002A&A...390..155D}. These results showed that brighter novae tend to evolve faster in the optical and exhibit larger ejection velocities. The main driver behind these relationships was believed to be the WD mass \citep{1992ApJ...393..516L}.

The receding of the nova photosphere and the accompanying hardening of its emission peak ultimately give rise to a supersoft X-ray source (SSS) with an effective temperature of less than 100~eV and no emission above 1~keV \citep{1998A&A...332..199P}. The SSS is powered by stable hydrogen burning within the part of the accreted envelope that was \textit{not} ejected during the outburst. This phase of the nova outburst is believed to occur generally \citep[e.g.][]{2006ApJS..167...59H}; however, it can only be observed when the ejected matter becomes optically thin to supersoft X-rays \citep{1989clno.conf...39S,2002AIPC..637..345K}.

In this paper (as in \mzk), we define the \textit{turn-on time of the SSS} (\tonk) as the time in days after the optical outburst at which the CN became visible in (soft) X-rays. Therefore, \ton is an observational parameter that depends on the detection limit of the specific X-ray observation. However, our homogeneous monitoring strategy (see Sect.\,\ref{sec:obs}) provided detection limits that were consistently sufficiently low to detect practically all realistic post-nova SSS phases, given that no strong additional absorption was present. This justifies using \ton in the statistical comparisons described in Sect.\,\ref{sec:discuss}.

The SSS phase is fuelled by the remaining hydrogen and its end indicates the cessation of the (stable) residual burning and the disappearance of the nova towards quiescence. The \textit{SSS turn-off time} (\toffk) is defined in some (theoretical) studies as the time the hydrogen burning switches off \citep[e.g.][]{2006ApJS..167...59H,2010ApJ...709..680H}. Here, as in \mbk, we define \toff observationally as a drop in the SSS luminosity below the X-ray detection limit. This limit only allows us to follow the decreasing X-ray emission of a nova in \m31 until it declines to a certain luminosity, usually 1--2 orders of magnitude below the peak. The SSS light curves of Galactic novae typically decline by several orders of magnitude and can be followed for longer \citep[e.g.][]{2011ApJ...727..124O}. Nevertheless, our \toff is expected to extend beyond the actual hydrogen burning switch-off by a certain amount of time. This time span is relatively short, because the luminosity decline happens quickly compared to the SSS phase duration \citep[e.g.][]{2013ApJ...777..136W}, but it will be non-negligible in some cases. This should be kept in mind when comparing our data to other (theoretical) studies.

Both time scales, \ton and \toffk, are measured in days after the optical outburst. \citet{2006ApJS..167...59H} found a ``universal decline law'', based on models describing free-free emission and an optically thick wind, and used it to study the multi-wavelength evolution of Galactic nova outbursts.

A nova outburst constitutes a surface eruption and ejection of a part of the accreted material. The WD itself is largely unaffected by this event \citep[it is believed that a certain extent of mixing between core and envelope is necessary; see e.g.][]{2010A&A...513L...5C}, and after a while, the resumed accretion can lead to another nova outburst. If two or more eruptions of the same nova have been observed within about a hundred years time, the object is called a recurrent nova (RN). This arbitrary, phenomenological definition arises from the current look-back time of modern professional astronomy. It is in debate if the WD grows in mass with every nova outburst and thus will ultimately exceed the Chandrasekhar mass, exploding as a type Ia supernova (SN~Ia) \citep[see e.g.][]{1996ApJ...473..240D,2005ApJ...623..398Y,2010Natur.463..924G,2012ApJ...756L...4H,2012BASI...40..393K,2013AJ....145..117S} 

If CNe were found to contribute significantly to the yet elusive group of SN~Ia progenitors, then another open question would gain additional importance: whether the CNe properties vary depending on the underlying stellar population. Motivated by early attempts to employ the light curves of CNe for extragalactic distance measurements \citep[see e.g.][]{1992ApJ...393..516L}, it became important to study any potential impact the characteristics of the host galaxy could have on these light curves and other observable parameters. The subject remains controversial. Some studies have suggested that the Hubble type of a galaxy has no significant influence on its luminosity-specific nova rate \citep[e.g.][]{1997ApJ...487L..45H,2000ApJ...530..193S}, while other studies argued in favour of nova rates dominated by old \citep[e.g.][]{1987ApJ...318..520C,1989AJ.....97.1622C} or young \citep[e.g.][]{1994A&A...286..786D,1997ApJ...481..127Y} stellar populations of the bulge or disk component of a galaxy, respectively.

The existence of two distinct nova populations was first suggested by \citet{1990LNP...369...34D} and \citet{1992A&A...266..232D} from observations of Galactic novae. Fast novae with $t_2 \leq 12$~d appeared to be associated mostly with the Galactic disk, while slower novae were concentrated primarily in the bulge or considerably above the Galactic plane. Later, \citet{1998ApJ...506..818D} studied spectroscopic nova populations based on the work of \citet{1992AJ....104..725W}, who had classified novae as either showing Fe II lines and low expansion velocities (``Fe~II novae'') or He and N lines, often with strong Ne lines, and high expansion velocities (``He/N novae''). \citet{1998ApJ...506..818D} reported that novae in the Galactic bulge mostly belong to the Fe~II type, whereas disk novae tend to exhibit He/N type characteristics. Recent photometric and spectroscopic observations of novae in the bulge-dominated galaxy \m31 and the disk dominated galaxy M\,33 are consistent with this result \citep{2011ApJ...734...12S,2012ApJ...752..156S}.

\begin{figure*}[!t]
  \resizebox{\hsize}{!}{\includegraphics[angle=0]{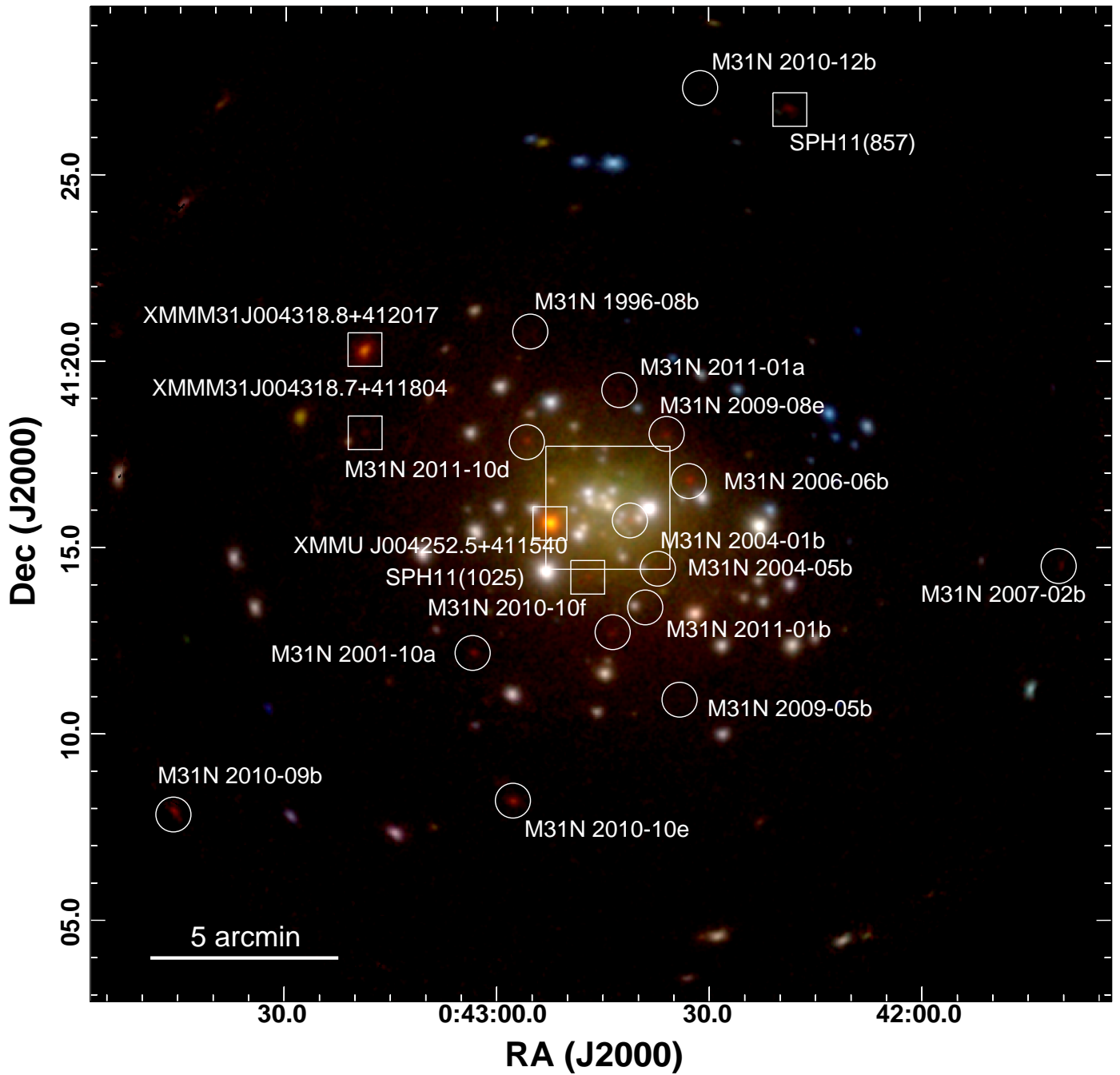}}
  \caption{Logarithmically scaled, three colour \xmm EPIC image of the central area of \m31 combining pn, MOS1, and MOS2 data for all 15 observations from Table\,\ref{tab:obs}. Red, green, and blue show the (0.2 -- 0.5) keV, (0.5 -- 1.0) keV and (1.0 -- 2.0) keV bands. Supersoft X-ray sources show up in red. The data in each colour band were binned in 2\arcsec x 2\arcsec pixels and smoothed using a Gaussian of FWHM 5\arcsec. The counterparts of optical novae detected in the outer regions of the fields, but not necessarily visible in this image, are marked with white circles. The non-nova SSSs detected in this work are marked by white boxes with SPH11(no.) referring to the source numbers in the catalogue of \citet{2011A&A...534A..55S}. The innermost $\sim 3\farcm3 \times 3\farcm3$ of \m31, as indicated by the large white box, suffer from source confusion in the \xmm data and the novae in this area are shown in a \chandra composite in Fig.\,\ref{fig:chandra}.}
  \vspace{1cm}
  \label{fig:xmm}
\end{figure*}

Our large neighbour galaxy, \m31 \citep[distance 780 kpc;][]{1998AJ....115.1916H,1998ApJ...503L.131S}, with its relatively low Galactic foreground extinction \citep[\nh $\sim 6.7$ \hcm{20},][]{1992ApJS...79...77S} is an obvious and excellent target for extragalactic nova surveys. Over the last century, beginning essentially with the seminal work of \citet{1929ApJ....69..103H}, more than 900 nova detections have been reported in \m31 \citep[937 candidate outbursts as of Jun 2013; see the online catalogue\footnote{http://www.mpe.mpg.de/$\sim$m31novae/opt/m31/index.php} of][hereafter \pzk]{2007A&A...465..375P}. Systematic discoveries of larger sets of \m31 novae in X-ray data began with \citet[][hereafter \pek]{2005A&A...442..879P}, who correlated X-ray catalogues from ROSAT, \xmmk, and \chandra with optical nova data and found that these objects constitute the major class of SSSs in \m31. Another successful archival study by \pz motivated a dedicated monitoring project, where the first results were reported in \mbk. For this project, the results of the most recent monitoring campaigns, their implications, and interpretations, are the subject of this work.

For an analysis of individual nova discoveries (X-ray spectra, light curves), several of which showed interesting features, we refer the reader to Sect.\,\ref{sec:results}. Those primarily interested in the discussion of the extended sample of \m31 novae with X-ray counterpart (parameter correlations, population studies) might find it useful to skip directly to Sect.\,\ref{sec:discuss}.

%
%
\section{Observations and data analysis}
\label{sec:obs}
%
This work is based on \xmm and \chandra observations of the central area of \m31 that were dedicated to the monitoring of SSS states of novae (PI: W. Pietsch). We report on the analysis of three observation campaigns carried out during Nov 2009 to Feb 2010, Nov 2010 to Mar 2011, and Nov 2011 to Mar 2012. Within these campaigns, the individual observations were separated by about ten days. The last campaign included a single \chandra observation at the beginning of Jun 2012. Additionally, we made use of two \xmm target of opportunity (ToO) observations of the \m31 disk nova M31N~2008-05d \citep{2012A&A...544A..44H} to constrain the X-ray parameters of a few objects (see Sect.\,\ref{sec:results}).

\begin{figure}[!t]
  \resizebox{\hsize}{!}{\includegraphics[angle=0]{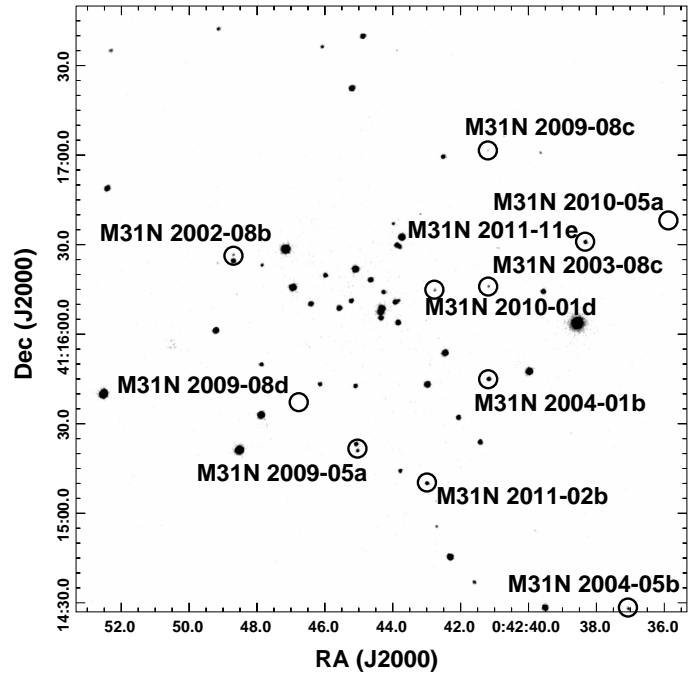}}
  \caption{Logarithmically scaled \chandra image of the innermost $3\farcm3 \times 3\farcm3$ of \m31 which combines all HRC-I observations analysed in this work (see Table\,\ref{tab:obs}). The images were not binned (HRC electronic pixel size = $0\farcs13$) but were smoothed with a Gaussian of FWHM $0\,\farcs5$. The X-ray counterparts of novae in the field, which are not all visible in this image, are marked with black circles.}
  \label{fig:chandra}
\end{figure}

In total, 39 individual monitoring observations have been obtained with an unscreened exposure of the order of 20~ks each. Their details are listed in Table\,\ref{tab:obs}. Hereafter, the three campaigns are named 2009/10, 2010/11, and 2011/12, respectively. In Fig.\,\ref{fig:xmm}, we show an \xmm image which was merged from all EPIC exposures and covers most of the area ($30\arcmin$ in diameter) of the monitoring. It includes all detected objects except of those in the innermost $3\farcm3 \times 3\farcm3$ of \m31 (shown in Fig.\,\ref{fig:chandra}). 

We used exactly the same instrumental setup as in \mbk: \xmm EPIC with pn in full frame mode and thin filter (MOS~1 and 2 with medium filter) and \chandra in HRC-I configuration. Although the HRC-I, a micro-channel plate detector, does not allow for the spectral fitting of the detected sources, it provides the largest field of view of all \chandra detectors, which was more important for establishing a dense monitoring of a relatively large area. The HRC-I also offers a good soft energy response. While \xmm provided a good spectral resolution and count rates, which allow for spectroscopic analysis (see Fig.\,\ref{fig:xmm}), the superb spatial resolution of \chandra did let us probe the innermost region around the \m31 core (see Fig.\,\ref{fig:chandra}) and discover many novae that fell victim to source confusion in the \xmm images.

Our data analysis had as its starting point the \xmm observation data files (ODF) and \chandra level 2 event files. These data were reprocessed using XMMSAS v11 \citep[\xmm Science Analysis System;][]{2004ASPC..314..759G}\footnote{http://xmm.esac.esa.int/external/xmm\_data\_analysis/} and CIAO v4.4 \citep[\chandra Interactive Analysis of Observations;][]{2006SPIE.6270E..60F}\footnote{http://cxc.harvard.edu/ciao/} with the calibration database (CALDB) version 4.4.7 and the latest calibration files. The analysis differed from the standard reduction chains and was described in detail for \xmm in \me and for \chandra in \citet{2013A&A...555A..65H}. These papers also described the creation of merged images that were used for each campaign to increase detection sensitivity.

Source lists, derived from XMMSAS \texttt{emldetect} and CIAO \texttt{wavdetect} output, were correlated against the most recent version of the MPE online \m31 optical nova catalogue. The correlation took into account the positional uncertainties from optical and X-ray detections. All luminosities given in this work were derived assuming a generic SSS spectrum with a black body temperature of 50~eV and Galactic foreground absorption (\nh $\sim 6.7$ \hcm{20}) and not on the basis of spectral analysis. Consequently, they are named ``equivalent luminosities'' (``$L_{\rm 50}$`` in source tables). We strongly advise not to quote these luminosities out of context. They should only be used to discuss relative changes of source flux within the monitoring campaigns. 

Spectral analysis was performed in XSPEC v12.7 \citep{1996ASPC..101...17A} using the single pixel events (and FLAG = 0) from the \xmm EPIC pn data because of the instrument's superior low-energy response. For a few objects, EPIC MOS data were also used, where we selected events with FLAG = 0 and PATTERN $\leq 12$. All spectral models use the T\"ubingen-Boulder ISM absorption model (\texttt{TBabs} in XSPEC) with the photoelectric absorption cross-sections from \citet{1992ApJ...400..699B} and ISM abundances from \citet{2000ApJ...542..914W}.

X-ray light curves were analysed after transforming the photon arrival times to the bary-centre of the solar system, using the XRONOS tasks of HEASARCs software package FTOOLS \citep{1995ASPC...77..367B}\footnote{http://heasarc.gsfc.nasa.gov/ftools/}. Additionally, we checked the \chandra light curves for indications of variability using the CIAO tool \texttt{glvary}, which applies the algorithm of \citet{1992ApJ...398..146G} to classify source variability. The statistical analysis in Sect.\,\ref{sec:discuss} was performed within the R software environment \citep{R_manual}.

Independently from the correlations with the optical nova catalogue, we used a hardness ratio criterion to search for SSSs in the \xmm data. The following formula, which was used in \mb to define hardness ratios (HR) and their errors (EHR), was adopted from \citet{2005A&A...434..483P}:

\begin{equation}
HR_i = \frac{B_{i+1} - B_i}{B_{i+1} + B_i} \; \mbox{and} \; EHR_{i} = 2 \frac{\sqrt{(B_{i+1}EB_i)^2 + (B_{i}EB_{i+1})^2}}{(B_{i+1} + B_i)^2}
\label{eqn:hardness}
\end{equation}

\noindent
for $i =$ 1,2, where $B_i$ and $EB_i$ denote count rates and corresponding errors in band $i$ as derived by \texttt{emldetect}. In \pek, these hardness ratios were used to classify sources as SSSs if they fulfilled the conditions $HR1 < 0.0$ and $HR2 - EHR2 < -0.4$. In this work, we used the same criteria mainly to find SSSs without a nova counterpart (see Sect.\,\ref{sec:res_sss}). The \xmm EPIC energy bands used here were (0.2 -- 0.5) keV, (0.5 -- 1) keV and (1 -- 2) keV ($i =$ 1 -- 3).

\section{Results}
\label{sec:results}
%
In total, we detected 24 X-ray counterparts of optical novae in this work. Seven of these sources were already X-ray active in the previous monitoring data presented in \mz (see Sect.\,\ref{sec:res_known} and Table\,\ref{tab:novae_old_lum}). Of those, all three SSS counterparts that were already active during the 2006/7 campaign (and before, see \mek) were observed to turn-off during 2009 -- 2012. We detected 17 novae in X-rays for the first time (see Sect.\,\ref{sec:res_new} and Table\,\ref{tab:novae_new_lum}). The positions of all objects are indicated in Figs.\,\ref{fig:xmm} and \,\ref{fig:chandra}.

Tables \ref{tab:novae_old_lum} and \ref{tab:novae_new_lum} contain X-ray measurements for all novae detected with a significance above $2\sigma$ (for \xmmk, in the (0.2--1.0) keV band, combining all EPIC instruments). Three novae. which were active SSSs at the end of the 2008/9 campaign and were described in \mz but were no longer detected, are listed in Table\,\ref{tab:novae_old_non} with their $3\sigma$ detection upper limits. In addition, we give $3\sigma$ upper limits for all novae, which had their optical outburst within a year before or during the individual campaigns in Tables\,\ref{tab:novae_ulim8}, \ref{tab:novae_ulim9} and \ref{tab:novae_ulim10}, respectively. In case of \xmmk, whenever possible, these upper limits were derived from the EPIC pn data, because this instrument has the highest sensitivity for soft X-rays. Five SSSs without a nova counterpart, which were already discussed in \mbk, have also been detected in these campaigns. They are indicated in Fig.\,\ref{fig:xmm} and summarised briefly in Sect.\,\ref{sec:res_sss} and Table\,\ref{tab:sss}.

Tables\,\ref{tab:novae_old_lum} -- \ref{tab:novae_ulim10} contain the following information: the name; coordinates; outburst date of the optical nova; the distance between optical and X-ray source (if detected); the X-ray observation and its time lag with respect to the optical outburst; the unabsorbed equivalent X-ray luminosity or its upper limit in the (0.2--1.0) keV band, which assumes a 50 eV black body spectrum with Galactic foreground absorption; and comments.
\subsection{X-ray counterparts of optical novae in \m31 known previously}
\label{sec:res_known}
Seven novae from previous campaigns were detected in this monitoring (see Table\,\ref{tab:novae_old_lum}). Among them, there were three sources that were already active prior to \me and have been found to experience the end of their exceptionally long SSS phases before the end of 2011/12. These sources were: M31N~1996-08b, M31N~2001-10a, and M31N~2004-05b. The SSS turn-off of another early nova, M31N~1997-11a, is discussed in Sect.\,\ref{sec:res_ulim}.

\textit{Nova M31N~1996-08b} was still active in 2009/10 but was not detected anymore in the following two campaigns. This allowed us to estimate an SSS turn-off time of about 13.8~yr ($5047$~d $\pm160$~d) after outburst, which is the longest ever observed for any \m31 nova. In the Galaxy, the current record holder is nova V723~Cas \citep[see][]{2008AJ....135.1328N}, which was still observed as a SSS for more than 14 years after outburst in 2009 \citep{2010AJ....139.1831S} and has not turned off as of May 2013 (Henze et al., in prep.). Another Galactic nova with a long SSS phase was GQ~Mus \citep[SSS turn-off after 10 years;][]{1995ApJ...438L..95S,2010AJ....139.1831S}.

\textit{Nova M31N~2001-10a} experienced its SSS turn-off between the 2010/11 and 2011/12 monitoring campaigns. By chance, the source was in the field of view of an \xmm ToO observation in Aug 2011 \citep[see][]{2012A&A...544A..44H}, where it was not detected. Therefore, the end of the SSS phase could be constrained to about 9.6~yr ($3511$~d $\pm78$~d) after its discovery in the optical. No significant differences have been found between the X-ray spectra extracted from the campaings of 2009/10, 2010/11, and those discussed in \mzk. Therefore, we use the best-fit values given in \mz in our analysis in Sect.\,\ref{sec:discuss}.

\textit{Nova M31N~2003-08c} was still active at the end of the monitoring, albeit only in merged observations at a very low luminosity (see Table\,\ref{tab:novae_old_lum}). Owing to its proximity to the \m31 centre, this source was only detected in the \chandra HRC-I observations.

\textit{Nova M31N~2004-01b} could still be detected in the \chandra HRC-I observations of all campaigns. In \mzk, no \xmm EPIC spectra had been extracted because of the position of the nova which was very close to the \m31 centre and to a bright persistent X-ray source nearby (see Figs.\,\ref{fig:xmm} and \ref{fig:chandra}). Here, the merged \xmm data from all three campaigns with the 2008/9 observations from \mz allowed us to extract a sufficient number of counts for spectral modelling. The merged spectrum can be fitted using an absorbed black body model with best-fit parameters \nh = $0.3^{+0.5}_{-0.2}$ \hcm{21} and $kT = 42^{+9}_{-12}$ eV. This classifies the source as an SSS. Fits to merged spectra extracted from individual campaigns resulted in best-fit parameters that agreed within the errors, which, however, were relatively large ($1\sigma$ values of (10 -- 20) eV). We took great care in extracting the background spectra used in the analysis but cannot rule out that nearby sources and a (soft) diffuse emission component might influence the derived source spectrum. This has to be taken into account when interpreting the spectral parameters.

\textit{Nova M31N~2004-05b} was detected again in the 2009/10 campaign, where it appeared to experience a significant drop in luminosity (see Table\,\ref{tab:novae_old_lum}). Since the source was not detected anymore in 2010/11, we might have observed its gradual SSS turn-off during the 2009/10 observations. Nevertheless, we took a conservative approach and estimated that the turn-off happened between the 2009/10 and 2010/11 campaigns.

\textit{Nova M31N~2006-06b} was active almost throughout the three campaigns but appeared to have turned off during the last \chandra observations in Feb till May 2012 (see Table\,\ref{tab:novae_old_lum}). We merged these last four HRC-I observations for better statistics but could not detect the object. A reasonably low upper limit is provided by observation 13279 in Feb 2012. Therefore, we assume that the SSS turn-off occurred between the last detection in \xmm observation 0674210401 and this upper limit (see Table\,\ref{tab:novae_old_lum}). Comparing the best-fit black body temperatures for the combined X-ray spectra extracted from the 2009/10 and 2010/11 observations with those from \mz suggests a cooling of the source, but the errors are large and the difference is not significant (kT $= 37^{+17}_{-15}$~eV $\rightarrow 20^{+12}_{-10}$ eV). Therefore, we merged all existing spectra and derived best-fit values of \nh = ($0.9\pm0.3$) \hcm{21} and $kT = 28^{+6}_{-5}$ eV, which have errors that are a factor $\sim$ 2 smaller than for the previous estimate in \mzk.

\textit{Nova M31N~2007-02b} was still detected during the 2009/10 campaign but had turned off by the time of the 2010/11 observations. As in \mzk, this source was located at large off-axis angles and only detected in those \xmm pointings, where the roll angle allowed it to be in the EPIC field of view. We fitted the combined \xmm X-ray spectrum, based on those observations where the source was detected, using an absorbed black body model. The resulting best-fit parameters were compatible within the errors with the values derived by \mz based on two observing campaigns. Therefore, we fitted the old and new spectra simultaneously to arrive at new best-fit parameters of $kT = 32^{+12}_{-9}$ eV and \nh = ($2.0^{+1.2}_{-0.9}$) \hcm{21}.

\subsection{X-ray counterparts of optical novae in \m31 discovered in this work}
\label{sec:res_new}
We discovered 17 new X-ray counterparts of \m31 novae, which are described below. Their (equivalent-luminosity) light curves can be found in Table\,\ref{tab:novae_new_lum}.

\subsubsection{M31N~2002-08b}
The optical nova was discovered by \citet{2012A&A...537A..43L} on 2002-08-26.56 UT. The optical light curves shown by \citet{2012A&A...537A..43L} suggest a relatively slow decline.

A faint X-ray counterpart was first detected in the \chandra observations of 2009/10. This source was located within the inner 2\arcmin\, of \m31, which is close to a known persistent source (see Fig.\,\ref{fig:chandra}). Therefore, \xmm was not able to resolve it and only \chandra data were available to study its evolution. Although the object was detected in several individual \chandra observations, we only quote the more reliable detections based on merged data in Table\,\ref{tab:novae_new_lum}.

We revisited the earlier observational data from \mz and confirmed that the X-ray source was not significantly detected in the 2008/9 campaign. There was, however, a possible source visible to the eye in the merged data of that campaign. This candidate object was below the detection threshold. Its position was close to the detections of the nova in the later data. By carefully comparing the suspected source with the actual detections of the current campaign, we found a positional offset, which caused us to conclude that this dubious source, whether real or merely a result of background fluctuations, was not identical with the nova. Its presence probably influenced the upper limit given in Table\,\ref{tab:novae_new_lum} for the 2008/9 campaign (''mrg2``), which has a larger value than the value for the merged data from the 2007/8 campaign (included as ''mrg1`` in Table\,\ref{tab:novae_new_lum}) where there were no suspicious sources visible.

As a result of this investigation, we estimated that the nova experienced the beginning of its SSS phase between the 2008/9 and 2009/10 monitoring campaigns. The source appeared to be still active at the end of the 2011/12 monitoring.

\subsubsection{M31N~2009-05a}
The optical nova was discovered by K.~Hornoch\footnote{http://www.cbat.eps.harvard.edu/CBAT\_M31.html\#2009-05a} on 2009-05-17.043 UT and confirmed in H$\alpha$ data by \citet{2009ATel.2147....1P}. A relatively faint X-ray counterpart was first discovered in the 2010/11 campaign and remained active until the end of the monitoring (see Table\,\ref{tab:novae_new_lum}). Owing to its position near a persistent X-ray source, this object was only detected in the \chandra observations.

The source was not detected in X-rays until the second observation of the 2010/11 campaign. It always remained close to the detection threshold (see Table\,\ref{tab:novae_new_lum}). We assume that its SSS phase began between the 2009/10 and 2010/11 campaigns (see Table\,\ref{tab:novae_new_lum}). The source was still detected at the end of the 2011/12 monitoring.

\subsubsection{M31N~2009-05b}
The optical nova was discovered by K.~Hornoch\footnote{http://www.cbat.eps.harvard.edu/CBAT\_M31.html\#2009-05b} on 2009-05-17.043 UT and confirmed by \citet{2009ATel.2147....1P} in H$\alpha$ observations. A faint X-ray counterpart was detected in the first \chandra observation of 2009/10 174~d after discovery. No previous X-ray observations of sufficient depth were available to constrain the SSS turn-on time further. We estimated that the SSS phase ended between the first and second campaign.

Based on the merged \xmm data of 2009/10, we fitted the \xmm EPIC pn spectrum of the source with an absorbed black body model, resulting in best-fit parameters $kT = 30^{+26}_{-22}$ eV and \nh = ($0.4^{+2.3}_{-0.4}$) \hcm{21}. Despite the relatively large errors, owing to a low-count spectrum, this source can be clearly classified as an SSS.

\subsubsection{M31N~2009-08c}
The optical nova candidate was discovered by K.~Hornoch\footnote{http://www.cbat.eps.harvard.edu/CBAT\_M31.html\#2009-08c} on 2009-08-12.423 UT \citep[see also][]{2009ATel.2165....1H}. The optical light curve appeared to evolve relatively slowly with a decline of only $\sim1.5$~mag by Sep 9 2009 28~d after discovery \citep[see][]{2009ATel.2213....1M}. An X-ray counterpart was clearly detected in the last observation of the 2009/10 campaign. No source was visible in the previous \chandra observation, leading to a well constrained SSS turn-on time. The source was still faintly detected in the first two pointings of the 2010/11 monitoring but its luminosity declined steadily. We estimated that the SSS turn-off happened between the last \chandra detection in observation 12111 and observation 12114, which provided a sufficiently low upper limit for us to reason that the source had turned off (see Table\,\ref{tab:novae_new_lum}).

\subsubsection{M31N~2009-08d}
The optical nova was discovered by K.~Hornoch\footnote{http://www.cbat.eps.harvard.edu/CBAT\_M31.html\#2009-08d} on 2009-08-12.423 UT \citep[see also][]{2009ATel.2165....1H}. \citet{2011ApJ...734...12S} found a moderately fast decline \citep[in the system of][]{1964gano.book.....P} for the optical light curve of $t_2 = (36\pm5)$~d in the $R$ band. The object was classified as an Fe~II nova in the system of \citet{1992AJ....104..725W} by \citet{2009ATel.2171....1D}, who gave an H$\alpha$ FWHM of 1300 km s$^{-1}$. A faint X-ray counterpart was found in the merged \chandra HRC-I data of the 2009/10 monitoring at the detection limit. Nothing was found at this position in the merged observations of 2010/11.

\subsubsection{M31N~2009-08e}
The optical nova was discovered independently by K.~Nishiyama \& F.~Kabashima\footnote{http://www.cbat.eps.harvard.edu/CBAT\_M31.html\#2009-08e} and \citet{2009ATel.2176....1O} on 2009-08-25.6293 UT. A very slowly declining light curve ($t_2 = 121$~d $\pm8$~d) was reported by \citet{2011ApJ...734...12S} based on $R$ band data. Observations by the Palomar Transient Factory \citep[PTF;][]{2009PASP..121.1395L} confirmed the slow decline in the optical \citep{2012ApJ...752..133C}. The nova was confirmed spectroscopically by \citet{2009ATel.2213....1M} as an Fe~II nova with an H$\alpha$ FWHM of 1230 km s$^{-1}$. It was detected as an ultraviolet (UV) source in \swift observations at about 58~d after discovery \citep{2009ATel.2274....1H}.

An X-ray counterpart was found as a faint source in the last \chandra observation of 2009/10. After being detected throughout the 2010/11 campaign, the object had disappeared in 2011/12. We estimated that the SSS turn-on happened between the last \xmm observation of 2009/10 and the first \chandra detection. The combined X-ray spectrum extracted from the \xmm data of 2010/11 (about 150 counts) could be fitted by a black body with best-fit $kT = 23^{+21}_{-14}$ eV and \nh = ($1.9\pm1.6$) \hcm{21}. This classifies the source as a SSS. 

\subsubsection{M31N~2010-01d}
The nova candidate was discovered close to the \m31 centre by \citet{2010ATel.2435....1P} in the \xmm optical monitor (OM) UV data taken on 2010-01-15.52 UT. Optical $R$ band detections of the object were reported by \citet{2010CBET.2187....3H} and showed that its light curve had declined relatively fast by about two magnitudes within 15 days after the first detection on 2010-01-16.795 UT. However, the optical peak was not well constrained with the closest non-detection on 2010-01-02.799 UT, so that the decay from maximum might have been faster.

Owing to its position near the \m31 centre, an X-ray counterpart was only detected in \chandra data. The source was visible from the first observation of the 2011/12 monitoring to the last \chandra observation in May 2012. Therefore, we can only give a lower limit for its SSS turn-off time.

\subsubsection{M31N~2010-05a}
The optical nova was discovered by \citet{2010CBET.2305....1H} on 2010-05-28.062 UT and spectroscopically confirmed as an Fe~II nova \citep{2010CBET.2319....1H}. It was independently discovered by \citet{2010CBET.2305....2N}. The object evolved slowly with a $t_2$ time of about 53~d \citep[see measurements given in][]{2010CBET.2411....1H}. \citet{2010ATel.2964....1P} reported H$\alpha$ detections of the nova in Oct 2010. The nova was also found in \swift UV observations during Jul and Aug 2010, displaying a slowly declining light curve \citep{2010ATel.2787....1H}. A very faint X-ray counterpart was only detected in the merged \chandra data of 2010/11. Nothing was found in the 2011/12 campaign (see Table\,\ref{tab:novae_new_lum}).

\subsubsection{M31N~2010-09b}
The optical nova was discovered by \citet{2010CBET.2472....1N}. \citet{2010ATel.2896....1P} confirmed the discovery and tightly constrained the time of outburst using a pre-discovery detection on 2010-09-30.412 UT and an upper limit on 2010-09-29.958 UT.  \citet{2012ApJ...752..133C} published a well sampled PTF $R$ band light curve and reported a fast rise (within two days) and decay ($t_2 = 10$~d). Two optical spectra have been obtained for this nova. The first was taken within a day of discovery, on 2010-10-01.39 UT, showing features of an Fe~II nova with an H$\alpha$ FWHM of 1300 km s$^{-1}$ \citep{2010ATel.2898....1S}. The second spectroscopic observation by \citet{2010ATel.2909....1S}, which was five days later on 2010-10-06.40 UT, revealed a significantly evolved spectrum with broader Balmer lines (H$\alpha$ FWHM of 3600 km s$^{-1}$). Although the latter spectrum showed some resemblance to those of hybrid novae \citet{2010ATel.2909....1S} reported that the initial classification as an Fe~II nova was confirmed.

An X-ray counterpart was detected in \xmm observations of 2010/11 near the edge of the field of view. Due to its large off-axis angles in the \chandra observations, the source was only detected in the merged data of this campaign, thereby not allowing us to put additional constraints on the SSS turn-on or turn-off time scales. M31N~2010-09b is likely to belong to the disk nova population of \m31; members of which were severely under-represented in the catalogue of \mzk. Therefore, we triggered an \xmm ToO observation in Aug 2011 to constrain the SSS turn-off time and/or X-ray spectrum of the source. In Table\,\ref{tab:novae_new_lum}, we show that the SSS phase of the nova had already ended by the time of the ToO \citep[which, however, detected another interesting disk nova: M31N~2008-05d, see][]{2012A&A...544A..44H}.

We fitted the \xmm EPIC pn and MOS spectra of the source simultaneously and derived best-fit black body parameters of $kT = (46\pm4)$ eV and \nh = ($4.6^{+1.0}_{-0.8}$) \hcm{21}. This classifies the source as an SSS. The resulting black body \nh is large and might lead to an underestimation of the source temperature, which therefore should be interpreted with care. Because of the location of the source at the edge of the EPIC field of view (see Fig.\,\ref{fig:xmm}) it was not always detected by all detectors in all observations, thereby limiting the number of source counts for spectroscopy.

\begin{figure}[t!]
  \resizebox{\hsize}{!}{\includegraphics[angle=0]{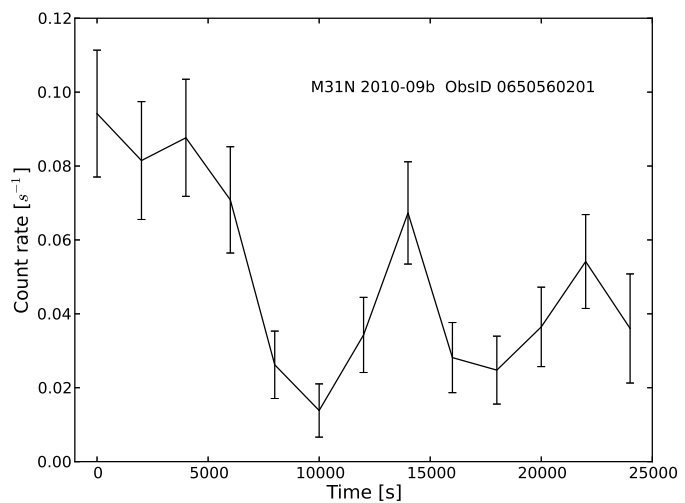}}
  \caption{\xmm EPIC pn $(0.2 - 1)$~keV light curve of nova M31N~2010-09b during observation 0650560201. The time is measured from the beginning of the exposure at UT 2010-12-26.43 UT with a 2000~s binning.}
  \label{fig:lc_n1009b}
\end{figure}

The EPIC pn $(0.2 - 1)$~keV light curve during \xmm observation 0650560201 (87~d after outburst) showed strong variability, which is plotted in Fig.\,\ref{fig:lc_n1009b}. On top of a declining trend, two broad dips can be identified with durations of about 6~ks each. There is no strong evidence of periodic behaviour. We did not find an energy dependence for the light curve nor a significant difference between the SSS spectra during and out of the dips. Nevertheless, these dips might indicate absorption effects as suspected for M31N~2008-05d \citep{2012A&A...544A..44H}, which arise possibly even from a re-establishing accretion disk \citep[see also][]{2012ApJ...745...43N}. In this case, M31N~2010-09b might be seen at high inclination. Of the other detections, only the MOS~2 data of ObsID 0650560401 suggested some (non significant) variability.

%

\subsubsection{M31N~2010-10e}
The optical nova was discovered by \citet{2010CBET.2573....1H} on 2010-10-30.703 UT. Its light curve declined fast. Comparing the measurements by \citet{2010CBET.2573....1H} to the PTF data reported in \citet{2012ApJ...752..133C} suggests that the first PTF detection (on 2010-11-01.158) took place close to maximum light and that $t_2 \lesssim 3$~d, which is the value that we adopted for our catalogue.

\citet{2010ATel.3001....1P} pointed out that the position of the object coincides (sub-arcsecond agreement) with three historical nova outbursts \citep[see][]{1973A&AS....9..347R,2012A&A...537A..43L}, which had already been discussed as multiple outbursts of a RN. However, due to the short gap of only five years between the first two outbursts, \citet{1989SvAL...15..382S} had suggested that the object might instead be a U~Geminorum dwarf nova system in the Galactic foreground. The question was solved when \citet{2010ATel.3006....1S} obtained an optical spectrum of the outburst, which clearly showed the object to be a He/N nova in \m31 with extremely broad Balmer emission lines (H$\alpha$ FWHM of 8100 km s$^{-1}$). Therefore, M31N~2010-10e was identified as the fourth recorded outburst of the RN M31N~1963-09c (discovered by \citet[][]{1973A&AS....9..347R}; the other two outbursts were M31N~1968-09a and M31N~2001-07b). Between its first two detected outbursts, this nova showed the shortest recurrence time ever observed (about five years) with the Galactic RN U~Sco in second place with about 10~yr \citep[see][]{2010AJ....140..925S}.

An X-ray counterpart was first discovered by \citet{2010ATel.3038....1P} in dedicated, high-cadence ToO monitoring observations with the \swift satellite only 15~d after outburst (with a non-detection two days earlier). \citet{2010ATel.3038....1P} also described a declining UV light curve. They classified the X-ray source as an SSS and reported indications for flux variability on time scales of hours. Due to the relatively large off-axis angle and the initial faintness of the source, our \chandra observations only detected the SSS on day 34 (see Table\,\ref{tab:novae_new_lum}). Fading rapidly, the source had disappeared during the \xmm monitoring. With the \swift observations, our monitoring provides accurate constraints on the duration of the SSS phase of this exceptional nova.

A simultaneous fit of the \xmm EPIC pn spectra extracted from the three observations, where the source was detected (see Table\,\ref{tab:novae_new_lum}) gave the following best-fit black body parameters: $kT = 61^{+6}_{-3}$ eV and \nh = ($3.1^{+0.4}_{-0.7}$) \hcm{21}. The relatively high SSS temperature supports the interpretation of a high-mass WD in the RN system.

\begin{figure}[t!]
  \resizebox{\hsize}{!}{\includegraphics[angle=270]{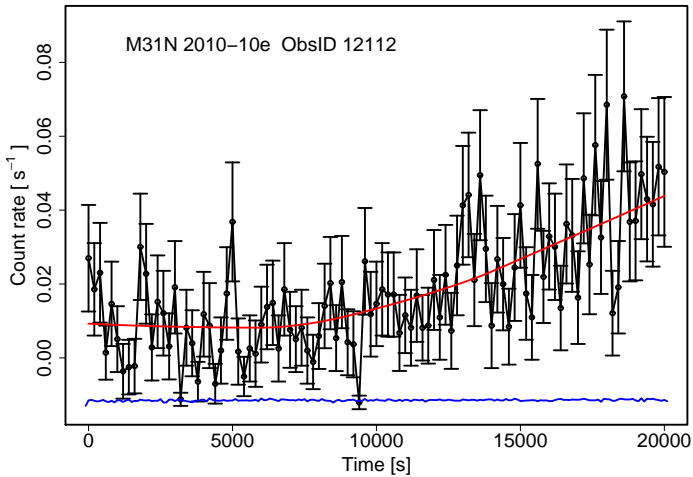}}
  \caption{\chandra HRC-I light curve of nova M31N~2010-10e during observation 12112. The time is measured from the beginning of the exposure at UT 2010-12-03.66 UT with a 200~s binning. The red curve is a smoothed fit to the light curve. The normalised and offset background light curve is shown in blue.}
  \label{fig:lc_n1010e}
\end{figure}

The light curve of M31N~2010-10e during the \chandra observation 12112 showed strong variability (\texttt{glvary} index of 9). The plot in Fig.\,\ref{fig:lc_n1010e} reveals that the count rate of the source gradually increased by a factor of about four (see smoothed red curve) after the first $(7-8)$~ks. This brightening can be attributed to the nova since the background light curve of the entire observation was very stable (blue curve). No variability was found in the short-term light curves of other observations of the source.

%

\subsubsection{M31N~2010-10f}
This nova in the \m31 globular cluster Bol~126 was first found in X-rays serendipitously by \citet{2010ATel.3013....1P} and is discussed in detail in \citet{2013A&A...549A.120H}. Its data relevant for the statistical analysis in Sect.\,\ref{sec:discuss} are given in Table\,\ref{tab:cat}.

\subsubsection{M31N~2010-12b}
The optical nova candidate was discovered by \citet{2010ATel.3076....1P} and several other observers independently \citep[][all in CBET \#2582]{2010CBET.2582....1K,2010CBET.2582....2P,2010CBET.2582....3N,2010CBET.2582....4S}. The first detection was on 2010-12-10.359. \citet{2012ApJ...752..133C} reported a very fast $t_2 = 3$~d based on a PTF monitoring light curve.

A faint X-ray counterpart with a very short turn-on time was found in \xmm observations of 2010/11. After being at the detection limit for the remaining \xmm observations of this campaign, the source was no longer detected in the last observations of 2010/11 (see Table\,\ref{tab:novae_new_lum}). However, owing to the large off-axis angles the source had in the \chandra observations of 2010/11, those measurements did not provide sufficient sensitivity to claim that the source had disappeared. Although the faintness of the source in the last \xmm ($2\sigma$) detection (day 56, see Table\,\ref{tab:novae_new_lum}) indicates that it might have faded below the on-axis detection limit not long thereafter, we take the more conservative approach and assume that the turn-off took place between 2010/11 and 2011/12.

We attempted modelling the X-ray spectrum extracted from \xmm observation 0650560301 (see Table\,\ref{tab:novae_new_lum}), which only had about 60 source counts (but considerably more than for the other observations). The best-fit parameters of a black body model show large errors, $kT = 39^{+21}_{-30}$ eV and \nh = ($0.9^{+8.6}_{-0.9}$) \hcm{21}, but nevertheless allow us to classify the source as SSS.

\subsubsection{M31N~2011-01a}
The optical nova was discovered independently by \citet{2011CBET.2631....1N,2011CBET.2631....2Y,2011CBET.2631....4H}; and \citet{2011CBET.2631....3S} with the first detection on 2011-01-07.39 UT. \citet{2011CBET.2631....6H} reported a noteworthy brightening to $R \sim 14.9$~mag on 2011-01-11.20 UT. The nova was detected in H$\alpha$ by \citet{2011ATel.3486....1H}. \citet{2011CBET.2631....7A} obtained an optical spectrum, showing the object to be an Fe~II nova with an H$\alpha$ FWHM of about 1300 km s$^{-1}$. After not being detected in 2010/11, a faint X-ray counterpart was found in the merged \xmm and \chandra data of the 2011/12 campaign (see Table\,\ref{tab:novae_new_lum}). 

\subsubsection{M31N~2011-01b}
The optical nova candidate was discovered by K.~Hornoch on 2011-01-16.725 UT. The object was announced on the CBAT transient objects confirmation page (TOCP)\footnote{http://www.cbat.eps.harvard.edu/unconf/tocp.html} under the designation PNV~J00423907+4113258. On this web site, also a confirming detection was posted by X.~Gao. 

The discovery of an X-ray counterpart was first announced by \citet{2011ATel.3441....1H} based on the \chandra data of 2010/11 with additional Swift observations in Jun 2011. \citet{2011ATel.3441....1H} classified the source as SSS, based on \swift XRT spectra, and gave a preliminary light curve, which showed the fast SSS turn-on (within a month after discovery) and covered the evolution until day 150 after outburst. Here, we extend the light curve until the end of the SSS phase.

Nothing was found at the position of the nova in the 2011/12 campaign. With a serendipitous detection in an \xmm ToO observation in Aug 2011 \citep[][]{2012A&A...544A..44H}, we were able to constrain the turn-off (see Table\,\ref{tab:novae_new_lum}). The ToO observation also provided an X-ray spectrum that could be fitted using a black body model with best-fit parameters $kT = 40^{+14}_{-20}$ eV and \nh = ($1.9^{+2.9}_{-0.9}$) \hcm{21}. We confirm the classification of the object as SSS by \citet{2011ATel.3441....1H} and the relatively high source temperature that was estimated based on \swift data.

\begin{figure}[t!]
  \resizebox{\hsize}{!}{\includegraphics[angle=0]{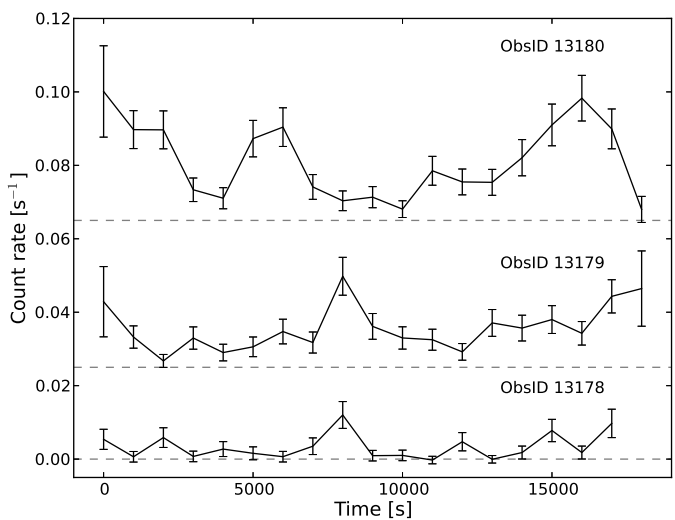}}
  \caption{\chandra HRC-I light curves of nova M31N~2011-01b during the three \chandra observations where it was detected. The time is measured from the beginning of each exposure at UT 2011-02-17.15 (ObsID 13178), 2011-02-27.25 (13179) and 2011-03-10.12 (13180). The upper two light curves include a count rate offset (see dashed grey zero levels) for better readability.}
  \label{fig:lc_n1101b}
\end{figure}

The short-term light curves of the nova during the three \chandra observations indicated variability (see Fig.\,\ref{fig:lc_n1101b}). The plots show that the variability and the average luminosity of the object increased with each observation (see also Table\,\ref{tab:novae_new_lum}). While the first two light curves had an \texttt{glvary} variability index of six (already to be interpreted as definitely variable) for the 13180 observation, this index increased to eight.

%

\subsubsection{M31N~2011-02b}
The optical nova candidate was discovered by K.~Hornoch on 2011-02-23.784 UT with an $R$ band magnitude of $(17.7\pm0.2)$~mag and announced on the CBAT TOCP (see above) as PNV~J00424296+4115104. Unfortunately, the optical outburst is not well confined because there only exists a non-detection upper limit on 2011-02-09.735 UT, which is 14~d before the first detection (K.~Hornoch, priv. comm.).

Constraining the outburst date would be of particular importance as a relatively bright X-ray counterpart was detected only 4~d after discovery in a \chandra observation of 2010/11 (see Table\,\ref{tab:novae_new_lum}). Such an extremely fast SSS turn-on would be unprecedented. Even when assuming that the outburst occurred immediately after the last non-detection, this would result in a turn-on time of $\lesssim18$~d (since the source could have been X-ray active even earlier), which places M31N~2011-02b among the SSSs with the fastest turn-on ever observed in \m31. However, novae that are fast SSSs usually also show a rapid decline of the optical light curve (see Sect.\,\ref{sec:discuss_corr}). Therefore,  it is unlikely that its optical brightness would have taken 14~d to decay by only two or three magnitudes to $R \sim 17.7$~mag unless M31N~2011-02b had been exceptionally bright in outburst ($R < 15.0$~mag). This suggests a significantly shorter SSS turn-on time than 18~d.

In the last \chandra observation of 2010/11, which occured ten days after the first detection, the luminosity of the X-ray counterpart had declined significantly (see Table.\,\ref{tab:novae_new_lum}). No X-ray source was detected at the position of the nova in the 2011/12 campaign.

\begin{figure}[t!]
  \resizebox{\hsize}{!}{\includegraphics[angle=270]{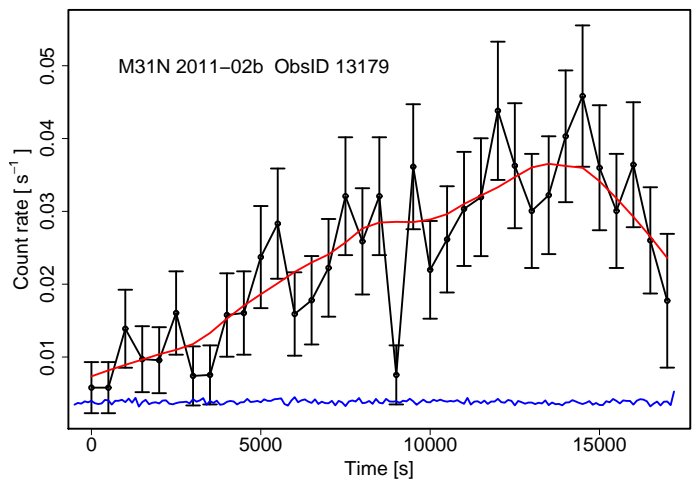}}
  \caption{\chandra HRC-I light curve of nova M31N~2011-02b during observation 13179. The time is measured from the beginning of the exposure at UT 2011-02-27.25 UT and binned in 500~s intervals. The red curve is a smoothed fit to the light curve. In blue, we show the background light curve in arbitrary scaling to illustrate the stability of the background count rate.}
  \label{fig:lc_n1102b}
\end{figure}

During the \chandra observation 13179, we observed strong short-term variability (\texttt{glvary} index of 9) while the source was brightest in X-rays (see Table\,\ref{tab:novae_new_lum}). As shown in Fig.\,\ref{fig:lc_n1102b}, the X-ray count rate increased by a factor of about five gradually during the observation. Towards the end of the exposure, a drop in luminosity is suggested in the last 2-3~ks. The X-ray background was quiet during the entire observation.

%

Unpublished optical photometry, which was made available to us by K. Hornoch, showed that the $R$ magnitude of the object had declined by $1.2$~mag within nine days after discovery. Assuming a steady decline rate from maximum, this suggests a \ttwo of $\sim 15$~d. However, the sparse optical data available did not agree with a smooth decline but indicated a brief re-brightening after an initial fall of $0.8$~mag in three days. This latter scenario therefore suggests a $\ettwo \sim 7$~d, which nevertheless is still too slow to explain the X-ray observations in the context of the population statistics presented in Table\,\ref{tab:cat} and discussed in Sect.\,\ref{sec:discuss_corr}.

To agree with the average behaviour displayed by most novae in the current catalogue, M31N~2011-02b would need to have had a $\ettwo \sim 1$~d, which would result in a very fast $\eton \sim 6$~d and $\etoff \sim 35$~d (see Eqns.\,\ref{eqn:ton_toff} and \ref{eqn:t2_ton}). However, such short time scales have not been observed before and consequently, any prediction would require an extension of the estimated parameter correlations beyond the ranges for which they were established (see Fig.\,\ref{fig:corr}). The analysis is further complicated by the fact that the SSS turn-off is not well constrained (see Table\,\ref{tab:novae_new_lum}) but might very well have happened only a few days after the last detection (extrapolating from the observed speed of the decline in luminosity).

It appears that M31N~2011-02b truly has been an exceptional nova, and its behaviour does not fit the correlations displayed by the parameters of the general sample (see e.g. Fig.\,\ref{fig:corr}). However, due to the uncertain outburst date, we deliberately refrain from estimating SSS turn-on and turn-off time scales and using them in the statistical analysis in Sect.\,\ref{sec:discuss}.

\subsubsection{M31N~2011-10d}
The optical nova was discovered by \citet{2011ATel.3693....1O} on 2011-10-19.715 UT. Follow-up detections were reported on the CBAT TOCP (see above). Initial optical spectroscopy was carried out at about 1.5~d after discovery by \citet{2011ATel.3699....1S} who confirmed the object as a nova in \m31. They classified it as an Fe~IIb (hybrid) nova and reported moderately broad Balmer lines with an H$\alpha$ FWHM of about 3300 km s$^{-1}$. A second optical spectrum was obtained by \citet{2011ATel.3725....1B} at about 7~d after discovery. They classified the nova as an Fe~II type and described strong emission lines with P~Cygni profiles and two distinct absorption components. \citet{2011ATel.3727....1S} obtained a third optical spectrum at about 10~d after discovery, which featured narrow emission lines (H$\alpha$ FWHM of 900 km s$^{-1}$) and emerging He~I emission. These properties led \citet{2011ATel.3727....1S} to revise their initial classification and conclude that the object might be an (unusual) Fe~II type nova.

A faint X-ray counterpart was detected in \chandra observation 13230 in Dec 2011 (see Table\,\ref{tab:novae_new_lum}). The source might have been visible (with less than $2\sigma$ significance) in the previous observation 12329 . After being active during all \xmm pointings of the 2011/12 campaign and appearing as a faint detection in Feb 2011 \chandra data, the source seems to have turned off by Mar 2012.

We fitted the spectrum extracted from the combined 2011/12 \xmm observations and derived best-fit black body parameters of $kT = 71^{+12}_{-13}$ eV and \nh = ($0.9\pm0.4$) \hcm{21}. Therefore, this source can be classified as an SSS.

\subsubsection{M31N~2011-11e}
\label{sec:res_1111e}
The optical nova was discovered by K.~Hornoch and J. Vrastil on 2011-11-19.704 UT and announced on the CBAT TOCP as PNV J00423831+4116313 (see there for confirmation detections). An optical spectrum was obtained by \citet{2011ATel.3778....1S}, who classified the object as a slowly-evolving Fe~II type nova with narrow, unresolved Balmer features (H$\alpha$ FWHM $< 500$ km s$^{-1}$).

An X-ray counterpart was first detected in a \chandra observation in Feb 2012. Owing to the position of the nova close to the \m31 centre, there was possible source confusion in some of the preceding \xmm observations, which resulted in not very reliable upper limits (see Table\,\ref{tab:novae_new_lum}). We took a conservative approach and assumed the \xmm observation 0674210501 as the last upper limit. The source was not detected anymore in the last \chandra pointing of the 2011/12 campaign.

%
\begin{figure}[t!]
  \resizebox{\hsize}{!}{\includegraphics[angle=0]{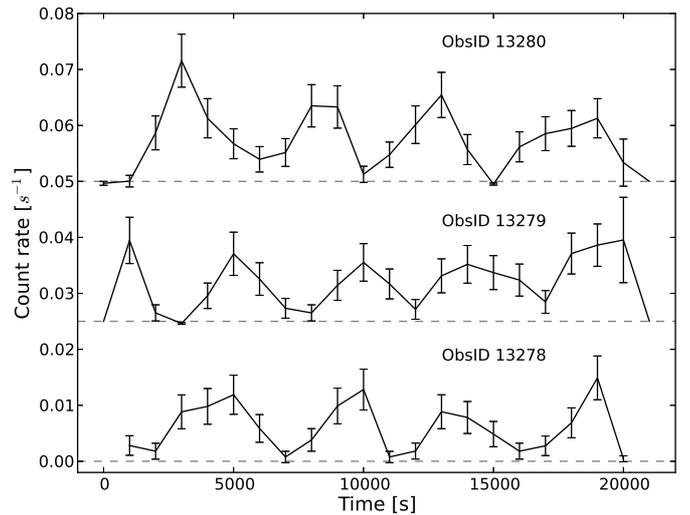}}
  \caption{\chandra HRC-I light curves of nova M31N~2011-11e during the three \chandra observations where it was detected. The time is measured from the beginning of each exposure at UT 2012-02-17.76 (ObsID 13278), 2012-02-28.26 (13279) and 2012-03-13.21 (13280). The upper two light curves include a count rate offset (see dashed grey lines) for better readability.}
  \label{fig:lc_n1111e}
\end{figure}

During all three \chandra observations where it was detected (see Table\,\ref{tab:novae_new_lum}), the \texttt{glvary} output for the source light curve strongly indicated variability. In Fig.\,\ref{fig:lc_n1111e}, we show the respective light curves. Fourier analysis suggested a periodic signal, and we used the XRONOS tool \texttt{efsearch} to determine the following best fit periods: 13278: $1.4\pm0.2$~h; 13279: $1.2\pm0.1$~h; and 13280: $1.4\pm0.2$~h. These results indicate that there were no significant changes in periodicity during the 20~d between the first and last detection. Assuming that the suspected frequency was stable, we estimated an average period of $1.3\pm0.1$~h.

Only one previous \m31 nova did show a similar suspected periodicity in its X-ray light curve: M31N~2006-04a with $P = 1.6\pm0.3$~h (in \mek). Note that the period found in the SSS flux of nova M31N~2007-12b, as discussed by \citet{2011A&A...531A..22P}, was considerably shorter ($\sim 1100$~s). While the result in \me was based on three possible cycles in a single \xmm observation, the observation of a consistent behaviour during three consecutive observations strongly increases the likelihood for an actual period for M31N~2011-11e.

In \mek, we discussed that periods longer than one hour in CV systems are most likely indicating the orbital period \citep[e.g.][]{2002AIPC..637....3W}, whereas pulsation periods are usually shorter \citep[see e.g.][2500 s pulsation period in nova V1494~Aql]{2003ApJ...584..448D}.

\subsection{Upper limits for non-detected X-ray emission of optical novae}
\label{sec:res_ulim}
Of the 10 novae that were still active at the end of \mzk, three were not detected in the present campaigns, namely M31N~1997-11a, M31N~2008-05a, and M31N~2008-06a. The two latter objects did turn-on during the 2008/9 campaign of \mzk; therefore, their relatively fast turn-off (about 400~d, see Table\,\ref{tab:cat}) is not surprising. Nova M31N~1997-11a displayed a remarkably long SSS phase. It showed a slow, gradual decline in X-ray luminosity during the three campaigns as discussed in \mbk and finally, twelve years after the optical outburst, it was not detected anymore in the 2009/10 observations. We estimate an SSS turn-off time of $\sim 11.5$ yr ($4207$~d $\pm182$ d), which makes M31N~1997-11a one of very few novae visible for more than a decade in X-rays. Upper limits for the three CNe are given in Table\,\ref{tab:novae_old_non}.

Additionally, we estimated upper limits for all novae not detected in X-rays and with optical outbursts between Oct 2008 and Feb 2010, between Oct 2009 and Mar 2011, and from Oct 2010 to May 2012 for the 2009/10, 2010/11, and 20011/12 campaigns, respectively. These values assume a confidence level of $3\sigma$ and are listed in Tables\,\ref{tab:novae_ulim8}, \ref{tab:novae_ulim9}, and \ref{tab:novae_ulim10}. We did not consider objects from the \pz MPE online catalogue, which were found not to be novae, but variable sources of other kinds.

\subsection{Non-nova supersoft sources}
\label{sec:res_sss}
We conducted a search for SSSs without a nova counterpart in the \xmm data based on the hardness ratio criterion described in Sect.\,\ref{sec:obs}. Five sources were found, which were all known before as SSSs or candidate SSSs in the \m31 catalogues of \citet{2005A&A...434..483P} and \citet{2008A&A...480..599S,2011A&A...534A..55S}. Table\,\ref{tab:sss} lists these objects and gives the corresponding source identifiers in the catalogue of \citet{2011A&A...534A..55S}. The five sources are identified in Fig.\,\ref{fig:xmm}.

\section{Novae with X-ray counterpart in \m31 - the updated catalogue}
\label{sec:cat}
%
In Table\,\ref{tab:cat}, we present the updated catalogue of all \m31 novae with a detected X-ray counterpart. This catalogue contains 79 objects; 38 of which were discovered in the dedicated monitoring project described here and in \mbk. It supersedes the catalogue published in \mzk. In addition to presenting the new novae described in Sect.\,\ref{sec:res_new}, the current version of the catalogue includes the improvements listed in the following paragraph.

For 32 novae detected by \xmmk, we carried out new, systematic fitting of their SSS spectra resulting in updated black body temperatures. We also included new optical spectroscopic and photometric information for many novae from the systematic study of \citet{2011ApJ...734...12S}. The catalogue contains two more new novae, which were not found in our monitoring: (i) the disk nova M31N~2008-05d, which was discovered by \citet{2012A&A...544A..44H} in \xmm ToO observations, and (ii) M31N~2012-05c, the SSS counterpart of which was found by \citet{2012ATel.4511....1H} in \xmm observations tracking the evolution of an ultraluminous X-ray transient in \m31 \citep[see][]{2013Natur.493..187M}.

As in \mzk, the catalogue in Table\,\ref{tab:cat} contains the following information: (a) For the optical nova, we include the name, date of outburst detection, maximum observed magnitude in a certain filter (not necessarily the peak magnitude of the nova), $t_2$ decay time in the R band, classification as belonging to the old/young stellar population (see Sect.\,\ref{sec:discuss_pop}), spectroscopic nova type in the classification scheme of \citet{1992AJ....104..725W}, and the maximum measured expansion velocity of the ejected envelope (half of the FWHM of the H$\alpha$ line). (b) For the X-ray counterpart, we include the turn-on and turn-off times, a flag for SSS classification, and the effective black body temperatures as inferred from the X-ray spectra. (c) It also includes derived parameters: the ejected and burned masses as computed according to Sect.\,\ref{sec:discuss_derived}; and (d) references. Note that not all parameters are known for all objects.

\section{Discussion}
\label{sec:discuss}
%
Throughout this section, the results reported in \mz constitute the starting points of our discussion. We expand them using new methods and approaches for the enlarged \m31 nova sample presented in Table\,\ref{tab:cat}.

\subsection{Derived nova parameters}
\label{sec:discuss_derived}
We estimated the amount of hydrogen mass ejected (\mejk) and burned (\mburnk) in each nova outburst based on the X-ray time scales. Using the same assumptions as in \mzk, we can describe the absorption generated by the expanding nova shell using the following formula:

\begin{equation}
N_{H} ({\rm cm}^{-2})= \emej/(\frac{4}{3}\pi \cdot m_H \cdot \evexp^{\;2} \cdot t^{2} \cdot {f}')\qquad.
\label{eqn:nh}
\end{equation}

Here, \mej is the ejected hydrogen mass, $m_H=1.673\times10^{-24}$ g the mass of the hydrogen atom and $f' \sim 2.4$ a geometric correction factor (defined in \me assuming a spherically symmetric shell based on \citet[][]{2002A&A...390..155D}). We assumed that the SSS turn-on at $t = \eton$ happens when \nh decreases to \ohcm{21}. Following \mzk, we used the (updated) correlation between the expansion velocity of the shell (\vexpk) and \ton, as modelled in Eqn.\,\ref{eqn:vexp_ton}, to eliminate \vexp from the model in favour of the much more frequently measured \tonk.

The ejected hydrogen masses, as given in Table\,\ref{tab:cat}, were computed using Eqn.\,\ref{eqn:nh}. They assume an inverse prediction of expansion velocities and their uncertainties based on Eqn.\,\ref{eqn:vexp_ton} with the respective turn-on times. The inverse prediction was performed using the R package \texttt{chemCal}, which implements the calculation of confidence intervals presented in \citet{Slutsky98a}. A note of caution: this prediction extends the relation between turn-on time and expansion velocity beyond the parameter ranges for which it was established in Eqn.\,\ref{eqn:vexp_ton}. This applies mainly to objects with very long SSS turn-on times ($\gtrsim 500$~d), where the uncertainties in their ejected masses might be underestimated. Measured expansion velocities were used in the case of novae for which they were known.

We estimated the mass of hydrogen burned in the WD atmosphere after the nova outbursts (\mburnk) as in \mbk. This formula contains the bolometric luminosity $L_{\mbox{bol}}$, SSS turn-off time \toffk, the hydrogen fraction of the burned material $X_H$, and the energy released by processing hydrogen $\epsilon=5.98\times10^{18}$ erg g$^{-1}$ \citep{2005A&A...439.1061S}.

Contrary to \mbk, here we included an estimate for the bolometric luminosity of an individual nova derived from its turn-off time. This was based on an approximation of the plateau luminosity during the constant bolometric luminosity phase by \citet{2005A&A...439.1061S}. We used their equation 2, which describes CO WDs, because these objects are expected to be intrinsically more frequent in our sample. There is, however, little difference between the CO model of \citet{2005A&A...439.1061S} and an ONe model with comparable metallicity (see their figure 6).

This luminosity approximation uses the WD mass, which we estimated following the models of \citet{2006ApJS..167...59H}. We took the model data given for WD mass and turn-off time in their Table 3 (\toff is called there $t_{\mbox{H-burning}}$), because the chemical composition of this model (''CO nova 2``) is similar to the CO WD scenario of \citet{2005A&A...439.1061S}. In particular, both models assume a hydrogen fraction of $X_H = 0.35$, which we consequently also used for our estimate. In Sect.\,\ref{sec:discuss_corr_context} below, we discuss a discrepancy between our data and a \ton vs \toff model by \citet{2006ApJS..167...59H}. However, this difference does not necessarily affect the \toff vs WD mass model, which only serves as a first order approximation at this stage. We parametrised the model data of \citet{2006ApJS..167...59H} using a broken power law in the $\log(\etoff) - M_{\mbox{WD}}$ plane. The resulting fit describes a dependency between luminosity and the logarithm of the turn-off time and is used in the following equation:

\begin{equation}
\emburn = L_{\mbox{\small{bol}}}\cdot \etoff / (X_H \cdot \epsilon)\qquad.
\label{eqn:mburn}
\end{equation}

\subsection{Correlations and Relationships between nova parameters}
\label{sec:discuss_corr}

\subsubsection{The correlations from \mz revisited}
\label{sec:discuss_corr_corr}

%
\begin{figure*}[t!]
  \resizebox{\hsize}{!}{\includegraphics[angle=0]{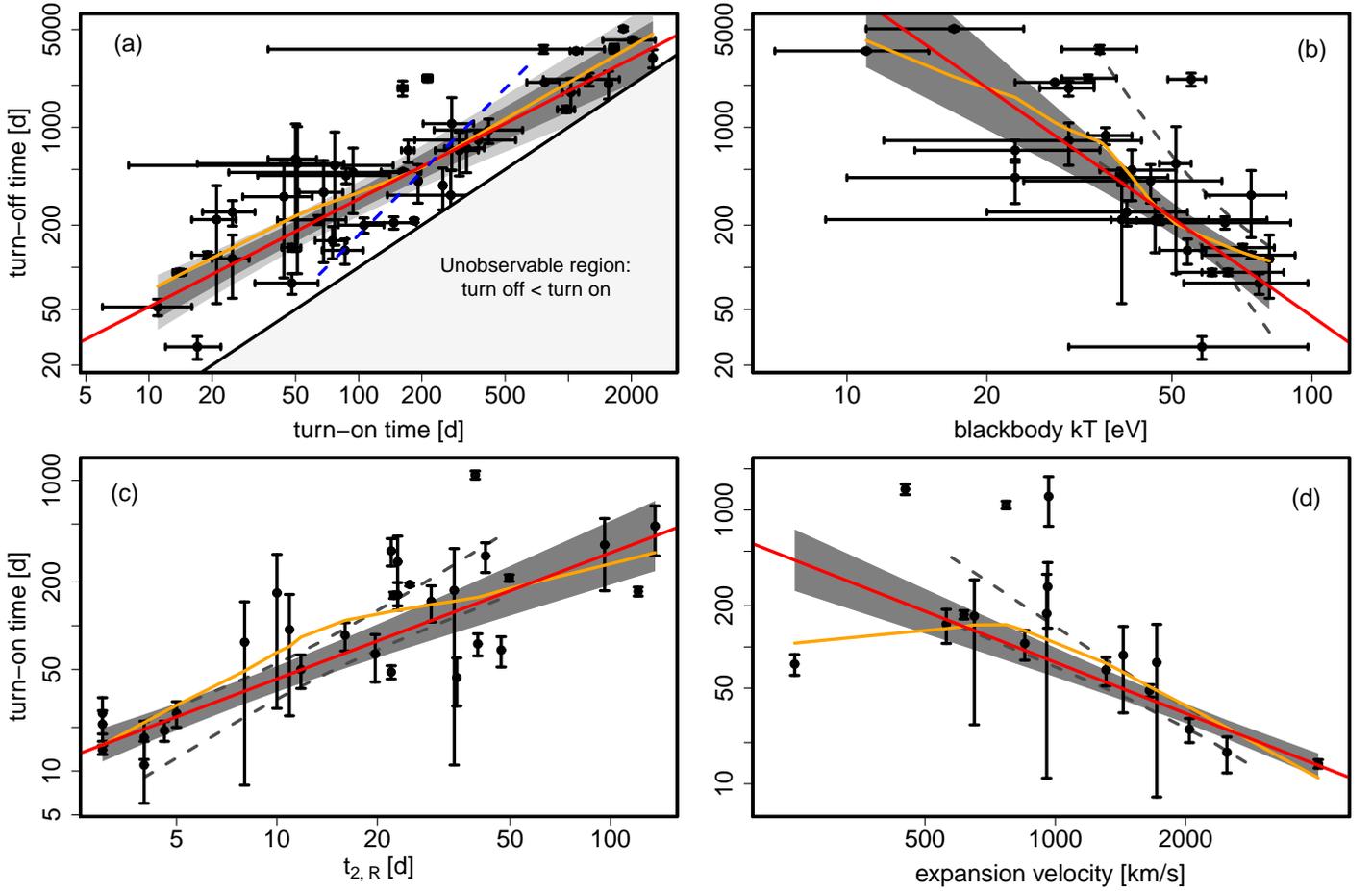}}
  \caption{Double-logarithmic plots of the updated correlations from \mzk. Data points and error bars are shown with a smooth fit (orange) for visualising, as well as a robust power law fit (red) with corresponding 95\% confidence regions (dark grey) for modelling. The correlations displayed are: (a) turn-on time versus turn-off time, (b) black body temperature (kT) in eV versus turn-off time, (c) optical decay time $t_{2,R}$ versus turn-on time, and (d) expansion velocity in km s$^{-1}$ versus turn-on time. All time scales are given in units of days after outburst. In panel (a), the light-grey shaded area around the best fit shows the 99.9\% confidence region; the blue dashed line indicates the relation found by \citet{2010ApJ...709..680H} for Galactic novae (see Sect.\,\ref{sec:discuss_corr_context}, and the lightly shaded area in the lower left corner visualises the ``unobservable region'' where the turn-off time occurs before the turn-on time. In panels (b)-(d), the grey dashed lines indicate the 95\% confidence regions from \mzk, which are based on an earlier version of the catalogue. For panel (a), we did not include these lines to avoid confusion and because there mainly is an improvement in accuracy but no change of slope.}
  \label{fig:corr}
\end{figure*}

In Fig.\,\ref{fig:corr}, we show updated double-logarithmic scatterplots for the four parameter correlations found in \mzk. Using the extended catalogue, all correlations are still present and we could reduce the uncertainty of the fit for most of them. Figure\,\ref{fig:corr} also compares the old and new confidence contours. We further included in each plot a smoothed representation of the scattered values to estimate the appropriateness of the power law fit. This smooth curve was estimated based on the LOWESS algorithm \citep{1981b_Cleveland}, which computes a robust locally weighted regression.

For this analysis, we only used those novae in Table\,\ref{tab:cat} for which the respective parameters were sufficiently well constrained, where their uncertainties were smaller than the values themselves. Furthermore, we estimated the best fits via a robust least squares regression method due to the presence of obvious outliers in all of the correlations. This fitting process (\texttt{rlm} in R's \texttt{MASS} package with default parameters) employed an M estimator \citep[see e.g.][]{2013pss2.book..445F}. Although there is no strong heteroscedacity left in the logarithmic variables, we continued to use weighted fits. All models in Fig.\,\ref{fig:corr} have been found to be stable using bootstrapping tests.

Furthermore, we confirmed the presence of strong correlations in the original, untransformed variables by computing the Spearman rank correlation coefficients for the four pairs of variables shown in Fig.\,\ref{fig:corr}. This is a non-parametric coefficient that does not assume a linear relation. Its absolute values for the correlations in Fig.\,\ref{fig:corr} a -- d are: $0.86$, $0.79$, $0.74$ and $0.65$. All four Spearman coefficients indicate correlations that are significant on the 99\% confidence level and beyond (p-values of 2\tpower{-14}, 3\tpower{-7}, 3\tpower{-6} and 6\tpower{-3}). In the following, we discuss the correlations in detail.

The correlation between the SSS turn-on (\tonk) and turn-off (\toffk) time is displayed in Fig.\,\ref{fig:corr}a. This plot contains the largest number of objects and shows the tightest correlation with a Pearson index of 0.86. Error bars for both time scales generally bridge the last detection and first post-SSS non detection with the parameter estimate at the midpoint in-between. Observing the turn-off of several novae with exceptionally long SSS phases provided us with stronger constraints towards the upper end of the relation. The best-fit power law index did not change with respect to \mz but the overall errors became slightly smaller: 

\begin{equation}
 t_{\mbox{off}} = 10^{(0.9\pm0.1)} \cdot t_{\mbox{on}}^{\quad(0.8\pm0.1)}\quad.
\label{eqn:ton_toff}
\end{equation}

The plot shows that the simple power law fit (red) follows the smoothed curve (orange) remarkably well over the entire parameter range. The remaining small deviations between the two curves might be explained by outliers (see below) or by a different behaviour of bulge and disk novae (see Sect.\,\ref{sec:discuss_pop}).

The light grey-shaded area in the lower right corner of the plot shows the ``unobservable'' region of $\etoff < \eton$. Note that objects in this region would not be forbidden, because physically \ton indicates the time at which the expanding nova envelope becomes transparent to soft X-rays and \toff is the time scale on which the H-burning ends. Recall that both time scales start at the optical outburst. It cannot be assumed a priori that there is no parameter configuration for which the H-burning ceases \textit{before} the ejected matter becomes optically thin to X-rays. Of course, these objects could never be observed. It remains an open question as to which extent the non-detection of some novae might be caused by such a self-absorption \citep[but see for instance][for a recent discussion on SSSs obscured by circumbinary material]{2013A&A...549A..32N}.

While this condition of not being able to populate the lower right corner of Fig.\,\ref{fig:corr}a restricts the parameter space, we argue that it does in no way presuppose the correlation. There is no obvious bias that would stop novae from entering the upper left corner of the plot nor hinder us from detecting them. These objects would necessarily be characterised by short \ton and long \toff times. This means that they should be visible for considerably longer times than any of the novae that are actually found as SSS. However, there is no detection of such objects even in the hundreds of kilo-seconds in our accumulated data. Therefore, we assume that the correlation we found is not caused by observational biases.

We found one possible source of distortions to the power law model: Our conservative strategy in estimating the \toff times for SSSs that disappeared between monitoring seasons. For those objects, we assumed a \toff at the midpoint between the campaigns. Very fast SSSs often turn-off during the campaign (however, see the discussion for M31N~2011-02b above). For long lasting SSSs, overestimating \toff by a few hundred days (i.e. the typical time between campaigns) does not make a big difference. However, adding such an offset to the turn-off times of medium fast novae could have a pronounced impact on the correlation.

Indeed, we see a group of objects above the power law fit with $\etoff \sim 200 - 1000$~d and large error bars in Fig.\,\ref{fig:corr}a. These are mainly novae from the early archival campaigns by \pe and \pz (e.g. M31N~2001-10f, M31N~2004-11e) for which the assumption of a \toff at the midpoint between observations might be an overestimate. This is a good example of how the identification of population trends can help to detect irregularities. However, without further knowledge of nova SSS light curves, which is a study beyond the scope of this paper, we have to consider the current individual estimates to be sufficiently cautious.

There are two novae clearly above the general data scatter: both M31N~2000-07a and M31N~2004-05b lie well outside the 99.9\% confidence region in Fig.\,\ref{fig:corr}a towards longer \toff values. Both sources are located on the upper end of the $kT$ vs \toff confidence contours in Fig.\,\ref{fig:corr}b, suggesting that their \toff might indeed be unusually long rather than their \ton being too short. The optical outburst of M31N~2004-05b is not well defined (discovery after visibility window re-opened), but the X-ray time scales for both objects are sufficiently long to be essentially independent of a few months shift of the nova outburst. Also, an earlier \toff around 2100~d (see Sect.\,\ref{sec:res_known}) would not alter the result significantly. Interestingly, both sources appeared to have experienced significant long-term variations in temperature and became significantly hotter towards the end of their SSS phase (see \pz and \mek). In \mzk, we speculated that prolonged SSS turn-off times could be explained by re-established accretion fuelling the H-burning on the WD beyond the expected duration.

The correlation between the black body temperature ($kT$) and \toff is the subject of Fig.\,\ref{fig:corr}b. Temperature error bars are $1\sigma$. Note that, due to a typing error, equation 3 in \mz (which modelled this relation for the earlier data) did not agree with the corresponding correlation plot. It should have been reading $\etoff = 10^{(8.6\pm1.3)} \cdot kT^{(-3.5\pm0.7)}$. We wish to thank J.~Osborne for pointing this mistake out to us. The relation stated in \mz did not take into account nova M31N~2007-12d (the fastest nova in the old and new correlation). The best-fit temperature of this nova does seem to deviate even more strongly in the updated plot (see Fig.\,\ref{fig:corr}b). All other objects agree reasonably well within the errors with the simple power law fit. This is also indicated by the way in which the power law approximates the smoothed curve. The updated model with a milder slope and reduced errors is given here:

\begin{equation}
 t_{\mbox{off}} = 10^{(6.3\pm0.5)} \cdot kT^{(-2.3\pm0.3)}\quad.
\label{eqn:kt_toff}
\end{equation}

The correlation of the R band light curve decay time by two magnitudes (\ttwok) and \ton is visualised in Fig.\,\ref{fig:corr}c. In comparison to \mzk, the slope of the power law fit and its overall uncertainties were significantly reduced. New data in both X-rays and the optical (from the photometric catalogue of \citet{2011ApJ...734...12S}) populated the fast end of the correlation. There remains considerable scatter, in particular towards the slower novae. Since the estimated \ttwo time depends on how close to the actual peak the maximum magnitude was measured and to a certain extent on the shape of the light curve, some of the scatter might arise from a necessarily non-continuous observational coverage. Nevertheless, the smoothed curve is reasonably well approximated by the best fit power law:

\begin{equation}
 \eton = 10^{(0.8\pm0.1)} \cdot \ettwo^{\;\;(0.9\pm0.1)}\quad.
\label{eqn:t2_ton}
\end{equation}

The correlation between the expansion velocity of the ejected envelope (\vexpk) as determined from optical spectra and \ton is shown in Fig.\,\ref{fig:corr}d. Here, the overall trend is still clearly visible, but the updated plot contains more scatter and significant outliers. This suggests a more complicated relation between the two parameters than was assumed in \mzk. Variances in density, composition, and geometrical shape of the ejected envelope are likely to create a scatter around this correlation. The smoothed curve is strongly attached to a single object with the lowest measured expansion velocity. This source might also influence the power law fit towards a milder slope. The current, robust model takes this data point into account and reads as follows:

\begin{equation}
 \eton = 10^{(5.6\pm0.5)} \cdot \evexp^{\quad(-1.2\pm0.1)}\quad.
\label{eqn:vexp_ton}
\end{equation}

In contrast to \mzk, we have included nova M31N~2003-08c in modelling this relation (the left-most object above the power law fit in Fig.\,\ref{fig:corr}d). Previously, the SSS turn-on time of this object was considered potentially unreliable. Now, the robust model is better capable of taking this nova into account without overestimating its influence.

The smoothed curve and, to a minor extend, the robust fit are influenced by the object with the lowest expansion velocity. This nova is M31~2011-11e, which is discussed in Sect.\,\ref{sec:res_1111e} as a potential high-inclination system, because of its apparently periodic X-ray light curve. With an asymmetric ejecta geometry, this inclination could lead to deviations from general population trends. Future extensions of the \m31 nova catalogue might be able to identify clusters of such objects and take them properly into account when modelling the average nova behaviour.

%
\begin{figure*}[t!]
  \resizebox{\hsize}{!}{\includegraphics[angle=0]{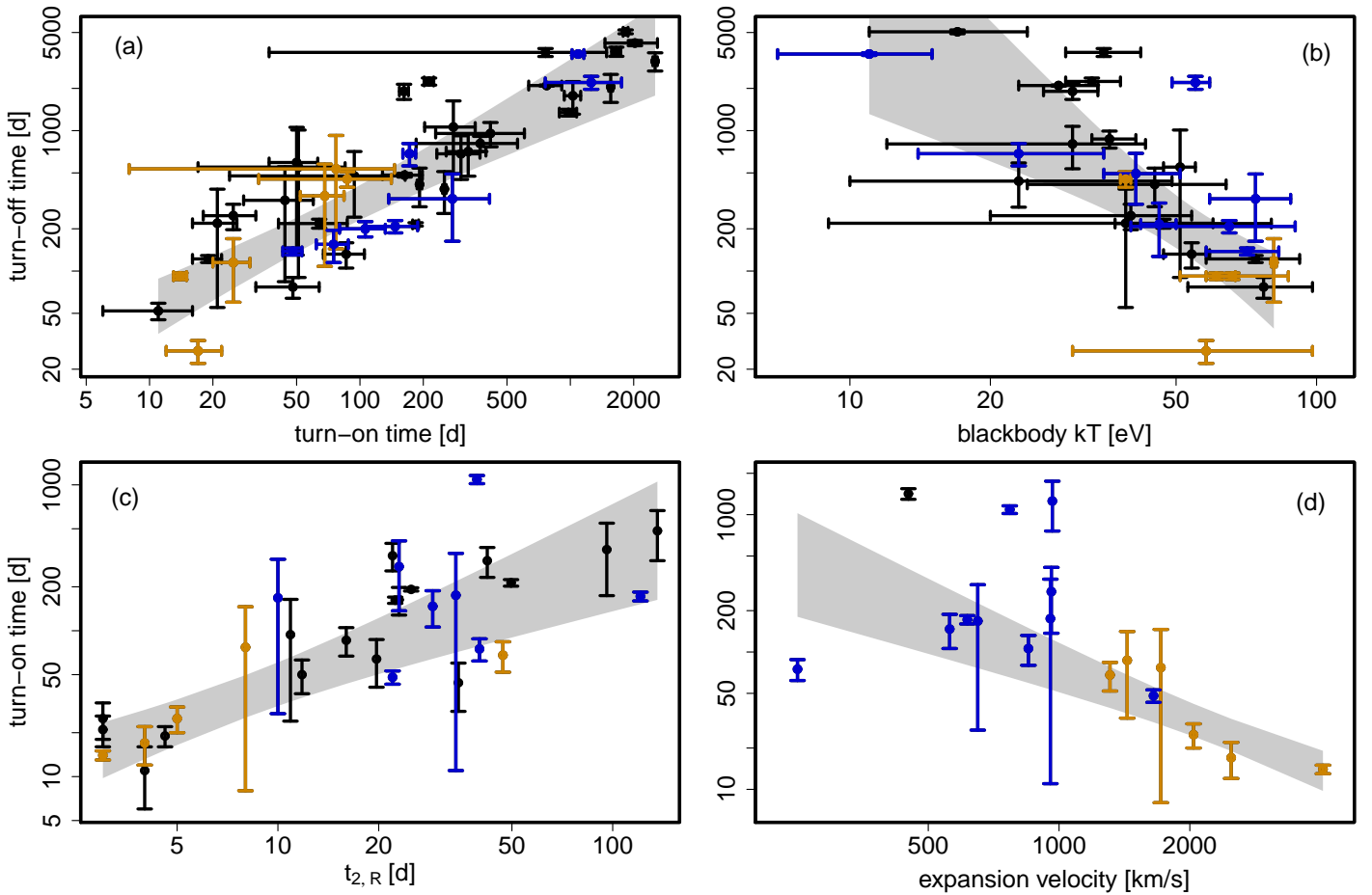}}
  \caption{Same as Fig.\,\ref{fig:corr} with colour-coded optical spectra classification (Fe~II=blue, He/N=orange) and grey best fit 99.9\% confidence regions. For black objects, there is no optical spectral classification.}
  \label{fig:corr_spec}
\end{figure*}

\subsubsection{The correlations in the context of literature results}
\label{sec:discuss_corr_context}
We compared the correlations and power law fits, as described in Sect.\,\ref{sec:discuss_corr_corr}, with results published in the literature. The similarities and differences we found are given here.

Concerning the relation between \ton and \toffk, Fig.\,\ref{fig:corr}a reveals that our fit for \m31 novae differs significantly from the theoretical prediction formula found by \citet{2010ApJ...709..680H} for the SSS phases of Galactic novae. The result from \citet{2010ApJ...709..680H}, as based on their optically thick wind theory and the ``universal decline law'' \citep{2006ApJS..167...59H}, is shown as a blue dashed line with a range of validity between $\sim 65$~d and $\sim 650$~d. This line lies outside the light grey-shaded confidence region (i.e. the 99.9\% level) of our power law fit. This comparison might be affected by systematic differences in the definition of the theoretical and observational time scales. For very faint SSSs, the detectability, and therefore, the turn-off time, depends on the detection limit of the specific observation or group of observations. In some cases, our observational estimate of \toffk, by the time the X-ray luminosity drops below the detection limit, could be longer than the theoretical \toff used by \citet{2010ApJ...709..680H}, which is the time of the actual hydrogen-burning switch-off. However, it is not clear whether these effects could have such a strong impact.

When we combine the relation found between SSS turn-on time and optical decay time ($\eton \propto \ettwo^{0.9}$; see Eqn.\,\ref{eqn:t2_ton}) with the estimate on the mass of the ejected envelope in Eqn.\,\ref{eqn:nh} ($\eton \propto \emej^3$), we find that $\emej \propto \ettwo^{0.3}$. This connection between the optical decay time and the ejected envelope mass, as derived from X-ray data, agrees well with a similar relationship based on optical data of Galactic novae \citep[][see their Figure 5 and the corresponding equation in their Section 6]{2002A&A...390..155D}.

Considering the relation between the two optical parameters (\ttwo and \vexpk), by combining Eqns.\,\ref{eqn:t2_ton} and \ref{eqn:vexp_ton}, produces a result that agrees within the errors with what was found by \citet{2011ApJ...734...12S} based on their spectroscopic and photometric survey of \m31. This indicates that the subset of \m31 novae with SSS phase does not seem to behave differently from the larger sample studied by these authors.

\citet{2001AJ....121.1126V} already suggested a correlation between SSS turn-off times (which they estimated using UV observations) and the \justttwo rate in Galactic CN data. These authors did not provide a fit to their correlation but it appears steeper than for \m31 novae, where we estimated a power law index $\sim 0.5$. However, it is not clear from their paper, which optical filter was used to observe the light curves that gave the \justttwo times and whether it was the same filter for all objects. We also note that these authors studied only ONe novae, which are expected to be a minor fraction in our sample. We prefer to use the \ton vs \ttwo relation because it is the cleaner and better correlation in our sample and provides an intuitive physical interpretation.

\citet{2003A&A...405..703G} published a relation for Galactic novae between \vexp and \toff with a power law index of $-2.1$ ($log(\etoff) = 9.65 - 2.1\cdot \log(\evexp)$ without uncertainties in the plot of their Figure 4). With $\eton \propto \etoff^{0.8}$ (Eqn.\,\ref{eqn:ton_toff}) and $\eton \propto \evexp^{-1.2}$ (Eqn.\,\ref{eqn:vexp_ton}), we found $\etoff \propto \evexp^{-1.5}$, which is considerably flatter. However, both relations might still be comparable within the (partly unknown) errors. Furthermore, Fig.\,\ref{fig:corr}(d) features some apparent outliers towards long turn-on times, which might suggest the need for a steeper power law model.

A very recent paper by \citet{2013ApJ...777..136W} simulated accreting WD properties using the Modules for Experiments in Stellar Astrophysics code \citep[\texttt{MESA},][]{2011ApJS..192....3P,2013ApJ...762....8D}. They also studied nova outbursts and found that the observed $kT$ - \toff data we published in \mz agreed with their simulations. However, we found in Sect.\,\ref{sec:discuss_corr_corr} that the slope of the corresponding power law model became flatter in the light of new data and might not be so similar to the results of \citet{2013ApJ...777..136W} for the slow novae. The simulation by \citet{2013ApJ...777..136W} also produced a relationship between WD mass and ejected mass, which is tightly connected to the SSS turn-on and turn-off times (their Figure 14). These results agreed with the \ton - \toff correlation from \mzk, which has been confirmed here.

Finally, \citet{2011ApJS..197...31S} recently published a comprehensive study of the SSS properties of Galactic novae, which was mainly based on \swift data. A detailed comparison between novae in \m31 and the Galaxy is beyond the scope of this work. This analysis will be the subject of an upcoming paper (Henze et. al., in prep.).

%
\begin{figure}[t!]
  \resizebox{\hsize}{!}{\includegraphics[angle=0]{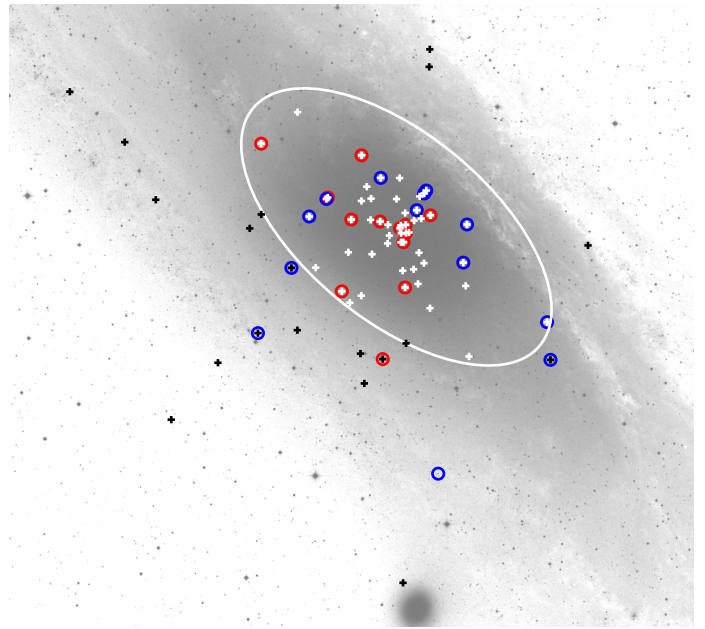}}
  \caption{Cut-out of \m31 image (from DSS2-R) overlaid with positions of bulge (white) and disk novae (black), where the large ellipse separates both populations (see Sect.\,\ref{sec:discuss_pop}). Circles mark novae with high-mass (blue) and low-mass (red) WDs, which were classified according to their SSS \ton times. See Sect.\,\ref{sec:discuss_pop} for the classification methods. Only four objects from Table\,\ref{tab:cat} are outside this image. North is up and east is left.}
  \label{fig:m31_bd}
\end{figure}

\subsubsection{Towards a multi-dimensional interpretation}
\label{sec:discuss_corr_pca}
In Fig.\,\ref{fig:corr}, it can be seen that the five parameters \tonk, \toff, $kT$, \ttwo and \vexp are all correlated with each other in a way such that novae that are fast in the optical also evolve quickly in X-rays and produce hot SSSs. The evidence that novae with short \ttwo times show fast ejection velocities had already been observed in large optical samples \citep[e.g.][see Sect.\,\ref{sec:discuss_corr_context} above]{2011ApJ...734...12S}. Our multi-wavelength data now suggests that novae might only populate a narrow strip in the five-parameter space of Fig.\,\ref{fig:corr}.

This is consistent with the result of a preliminary multi-dimensional analysis, which we describe in the following. We carried out a principal component analysis (PCA) on the X-ray parameters \tonk, \toffk, and $kT$. The PCA is an exploratory tool to reduce the parameter space of a multi-variate data set by finding new, uncorrelated variables (the so-called principal components; PCs) along axes of maximum variance in the original data. This process can be thought of geometrically, as a rotation that aligns the new axes with the spread of the data.

There are 33 novae for which the three X-ray parameters are known. The PCA input consisted of the log-transformed and standardised variables, thereby correcting for the different physical scales and the potential non-normality of the original parameter distributions (Shapiro-Wilk test: a normal distribution of the transformed variables cannot be excluded). The result shows that about 80\% of the total variance in the data could be attributed to the first PC (PC1). This indicates that there is a single hidden parameter, which is proportional to PC1, that dominates the behaviour of novae in the X-ray parameters.

These results have to be considered preliminary, because the sample size is still relatively small. A PCA relies on correlations, which could be spurious in a small sample, and although there seems to be no general consensus, the minimum recommend sample size appears to be around 50 objects. However, we found that the PCA results were stable under bootstrapping tests: $80 \pm 5$ percent of variance were attributed to PC1; the composition of which (i.e. the linear combinations of the original variables) was stable as well. We obviously cannot draw any conclusions beyond this first principal component but the clear dominance of PC1 indicates that most of the multi-parameter behaviour of our nova sample might be understood in surprisingly simple terms.

Nevertheless, this type of analysis clearly reveals the shortcomings of our data set. The relatively large number of objects in our catalogue easily hides that we only have measurements of a few parameters for many of them. We also considered a PCA of the full parameter space of Fig.\,\ref{fig:corr} but there were too few (only eleven) objects for which all five parameters had been measured. Additional data is necessary before confident conclusions can be drawn from such a multi-dimensional analysis.

If the initial impression of our preliminary analysis should be confirmed by future studies, what would this indicate? The correlations that are obvious in Fig.\,\ref{fig:corr} do not allow us to conclude that there are direct causal links between the various parameters. Indeed, it would be hard to understand how, for example, the SSS turn-on time could influence the turn-off time in Fig.\,\ref{fig:corr}a because the two time scales depend on different physical processes (see Sect.\,\ref{sec:discuss_derived}). The natural explanation for such a correlation is the existence of a third, ``hidden'' parameter (not directly measured in the data) that determines both of the correlated parameters.

Previous theoretical and observational studies provided several good candidates for this fundamental parameter, which are most prominently the WD mass \citep[e.g.][]{1992ApJ...393..516L,1995ApJ...452..704D,2002AIPC..637..443D,2006ApJS..167...59H}, the chemical composition of the WD \citep[e.g.][]{2005A&A...439.1061S,2006ApJS..167...59H}, the metallicity of the accreted material \citep[e.g.][]{2011ApJ...734...12S,2013ApJ...779...19K} or the accretion rate \citep[e.g.][]{1982ApJ...253..798N,2005ApJ...623..398Y}. All of these parameters appear to have a significant impact on the nova characteristics. This makes it difficult to reconcile all of the previous studies with our new results, which suggest that one of these parameters in the \m31 sample dominates the observable nova properties.

At this early stage, we resist the temptation to speculate which of the candidate parameters might dominate in our data, but it should be emphasised once again that the surprisingly small scatter around the \ton vs \toff relation in Fig.\,\ref{fig:corr}a indicates that the influences of other underlying characteristics, which are unrelated to the dominating parameter, appear to be minor in our \m31 nova sample. Note, however, that we present indications for differences between bulge and disk novae that might be related to a physical parameter of secondary importance in Sect.\,\ref{sec:discuss_pop}.

%
\begin{figure*}[t!]
  \resizebox{\hsize}{!}{\includegraphics[angle=0]{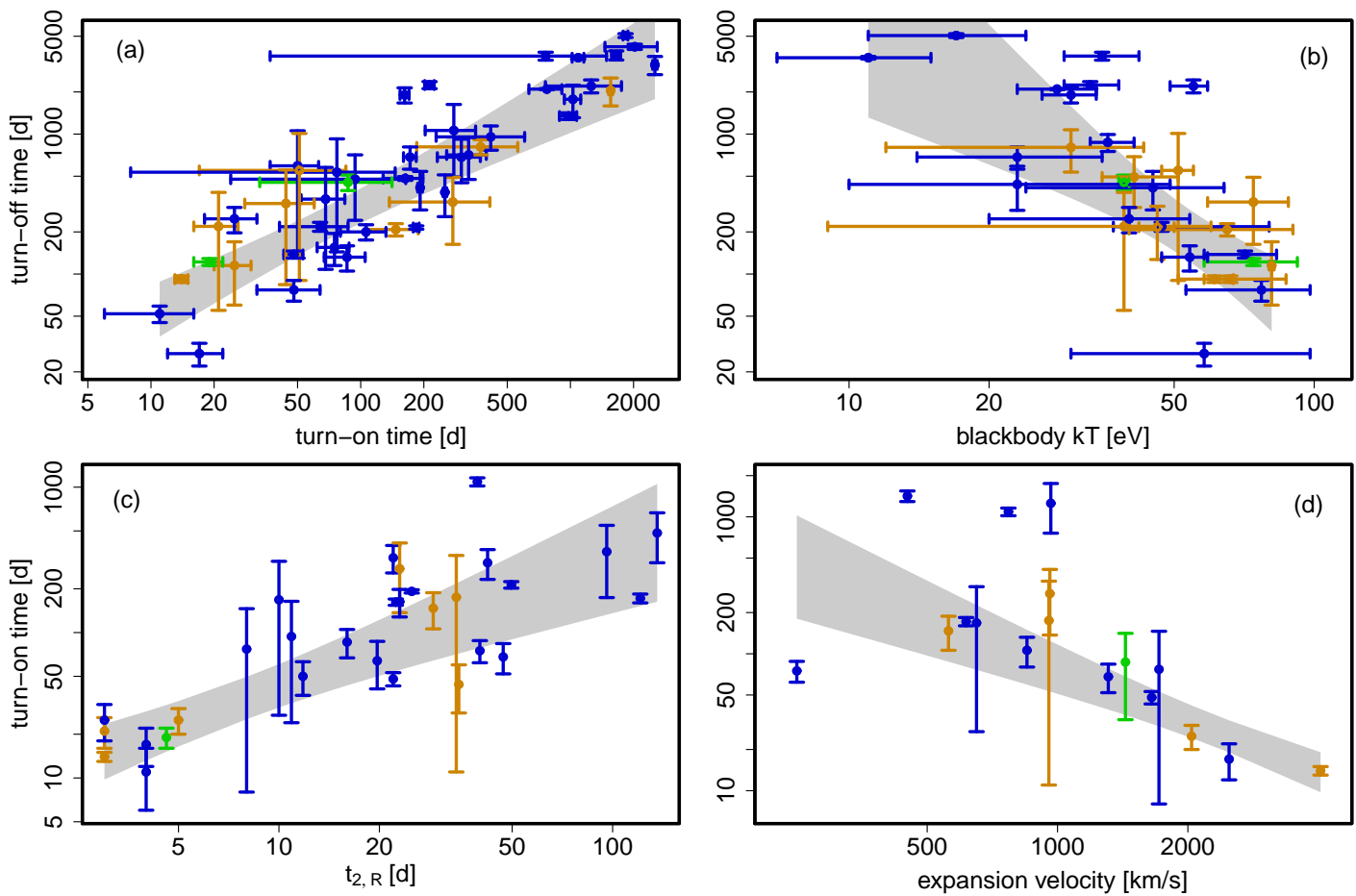}}
  \caption{Same as Fig.\,\ref{fig:corr} with colour-coded positional classification (bulge=blue, disk=orange, GC=green) and grey best-fit 99.9\% confidence regions.}
  \label{fig:corr_pos}
\end{figure*}

\subsection{Population estimates}
\label{sec:discuss_pop}
In this section, we discuss whether there are significant differences in the X-ray parameters of various sub-samples of novae. We distinguish between (i) Fe~II vs He/N novae \citep[according to their optical spectra in the system of][]{1992AJ....104..725W}, (ii) bulge vs disk novae, and (iii) novae with massive vs less massive WDs. The latter two groups of sub-samples are defined below.

In Fig.\,\ref{fig:corr_spec}, which is based on Fig.\,\ref{fig:corr}, we show how members of the two spectral classes He/N and Fe~II are distributed within the correlations, as discussed in Sect.\,\ref{sec:discuss_corr_corr}. Unsurprisingly, both spectral types separate strongly in their optical parameters \ttwo and particular \vexpk. The dichotomy in the latter case is one of the defining differences between those two classes \citep{1992AJ....104..725W}.

Nevertheless, Fig.\,\ref{fig:corr_spec}d underlines the finding that short SSS turn-on times are connected to He/N novae with high ejection velocities. \citet{1998ApJ...506..818D} reported that Galactic He/N novae tend to be associated with the disk stellar population, which is generally younger and more massive. Therefore, a short SSS turn-on should indicate a high WD mass. The two spectral classes of novae separate reasonably well in the X-ray parameter space as well, thus underlining the close interconnection between the behaviour of novae in both wavelength regimes. On average, He/N novae (orange in Fig.\,\ref{fig:corr_spec}) tend to be faster and hotter in X-rays. This supports the view that the two classes are related to fundamental parameters of the nova system.

For the classification of novae in bulge vs disk and high-mass vs low-mass subgroups, we used the same approach as in \mzk. The (projected) \m31 bulge was defined as an ellipse with a semi-major axis of 700\arcsec, an ellipticity of 0.5, and a position angle of $\sim 50\degr$ \citep[based on][]{2007ApJ...658L..91B}. Note that the two GC novae (M31N~2007-06b and M31N~2010-10f) have been excluded from this analysis because their environment is different from both bulge and disk \citep{2013A&A...549A.120H,2013ApJ...779...19K}. High mass WDs are defined as $M_{WD}\gtrsim1.2$\msun, which corresponds to $t_{on}\lesssim100$d, and low mass WDs as $M_{WD}\lesssim0.7$\msun, for $t_{on}\gtrsim500$d. See \mz for details. In Fig.\,\ref{fig:m31_bd}, the spatial distributions of the four sub-groups are shown, which are overlaid on an optical image of \m31 with a large ellipse indicating the bulge-disk boundary.

\begin{figure}[t!]
	\resizebox{\hsize}{!}{\includegraphics[angle=0]{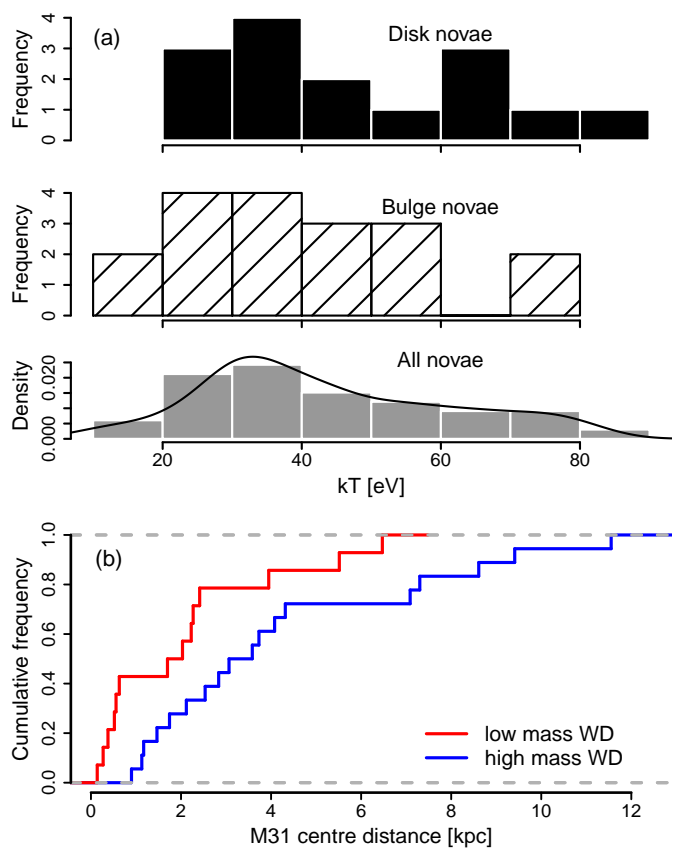}}
	\caption{Plot (a): Distribution of effective (black body) temperature $kT$ for disk novae (shaded/upper panel), bulge novae (black/middle panel), and the total sample (grey/lower panel), respectively. The upper two panels show frequencies and the lower one shows densities with an overlaid smooth density curve. Plot (b): Empirical cumulative density distribution functions of the inclination-corrected \m31-centric distances for novae with high mass (blue) and low mass (red) WDs. Distances are given in kpc, assuming a distance to \m31 of $780$ kpc and are not corrected for projection effects.}
	\label{fig:pop}
\end{figure}

In Fig.\,\ref{fig:corr_pos} we show how bulge vs disk novae arrange themselves in the five-parameter space of Fig.\,\ref{fig:corr}. Here, the picture is less clear than for the He/N vs Fe~II novae. While there appears to be clustering in the black body $kT$ parameter with disk novae that are on average hotter than bulge novae, the difference is not significant. This is visualised in Fig.\,\ref{fig:pop}a, which compares the $kT$ distributions for both sub-samples. While in \mzk, we found that the average black body temperatures of the two samples were significantly different on the 95\% confidence level; this result is not confirmed in the extended sample (using a Wilcoxon rank sum test). This might indicate that the earlier result was due to chance, which is not unlikely given our 95\% confidence criterion.

%

On the other hand, a related finding from \mz is confirmed in this work. This result focused on the distances from the \m31 centre of high-mass vs low-mass WDs, which are plotted in Fig\,\ref{fig:pop}b (using the same colours as for Fig\,\ref{fig:m31_bd}). The plot indicates a difference between the (empirical cumulative) distributions for the two samples, and a Wilcoxon test confirms the significance of this result on the (predefined) 95\% confidence level. With a p-value of 0.009, the significance even exceeds the 99\% confidence limit. Based on their recent spectroscopic and photometric survey of \m31 novae \citet{2011ApJ...734...12S} reported ``a weak dependence of speed class on position in \m31, with the spatial distribution of the fastest novae slightly more extended [...]''. Since optically fast novae also show a rapid SSS \ton (Eqn.\,\ref{eqn:t2_ton}), the method of \citet{2011ApJ...734...12S} is very similar to our comparison in Fig.\,\ref{fig:pop}b.

In view of these contrasting results, it remains puzzling that no compelling evidence can be found in favour of or against a dependence of the parameters of novae on their position in \m31. Clearly, the high inclination of \m31 \citep[$77.5\degr$; see e.g.][]{2007ApJ...658L..91B} leads to projection effects, which complicate an accurate positional classification for most novae. To overcome this difficulty, it would be necessary to specifically observe novae in the outer disk of \m31 for which an association with the bulge can be excluded. Presently, disk novae are still significantly under-represented in our catalogue.

\begin{figure}[t!]
  \resizebox{\hsize}{!}{\includegraphics[angle=0]{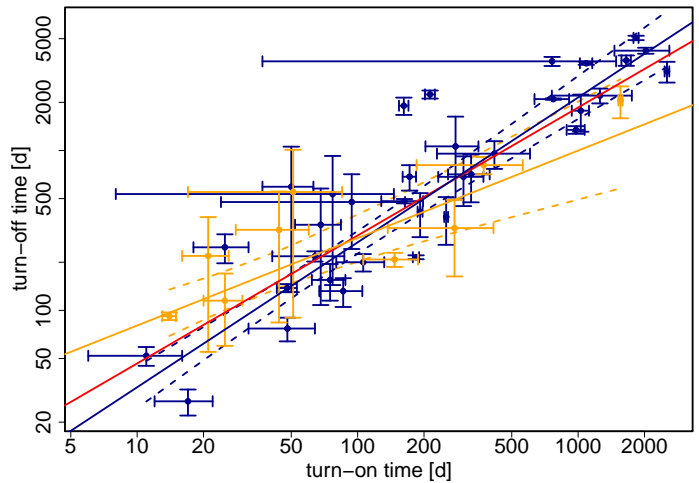}}
  \caption{Same as Fig.\,\ref{fig:corr}a, with different colours of symbols and best-fit lines for bulge (dark blue) and disk novae (orange). The two GC novae M31N~2007-06b and M31N~2010-10f have been excluded from this analysis. The dashed lines show the 95\% confidence limits associated with the fits. The overall best fit is shown in red.}
  \label{fig:ton_toff_bd}
\end{figure}

Furthermore, we have now found that bulge and disk novae are separated at a modestly significant level in the \ton vs \toff diagram, which is much more than in \mzk, where such a trend was merely suggested. Repeating the weighted robust linear regression (in log space) for \ton vs \toff (see Fig.\,\ref{fig:corr}a) separately for the bulge and disk novae, we found a significant difference. The model slope for bulge novae ($0.90\pm0.06$) turned out to be significantly steeper than for disk novae ($0.55\pm0.14$). These slopes agree with what was found in \mz but the uncertainties now have been reduced sufficiently for an analysis of variance to yield a significance on the 95\% confidence level (p-value of 0.016). The result is visualised in Fig.\,\ref{fig:ton_toff_bd}, and we found its (95\%) significance to be robust against the removal of outliers. Future studies should test this significance. Again, both GC novae were excluded in this analysis.

Such a difference in slopes for the \ton vs \toff relation in different populations is not predicted by current theoretical models. Could it be explained by observational biases that affect disk novae differently than bulge novae? Since a disk nova would not necessarily have been detected in our \m31 centre monitoring, which is in contrast to the majority of bulge novae, it could be possible for the less frequent and more irregular observational coverage to create a bias.

However, our sample only contained objects with measured SSS turn-off times and relatively well constrained errors in both time scales (see Sect.\,\ref{sec:discuss_corr}). This incidentally means that all disk novae considered in Fig.\,\ref{fig:ton_toff_bd} are located within the field of view of the central monitoring. Also, the average (median) uncertainties for the SSS turn-on and turn-off time scales were comparable for the bulge and disk sub samples.

Another possible way of distorting the measured \ton and \toff times of disk novae with respect to bulge novae is via absorption within \m31. On average, objects in the disk suffer a considerably higher extinction than those in the bulge \citep[see e.g.][]{2009A&A...507..283M}. The impact on the observed SSS duration is difficult to quantify without a detailed knowledge of SSS light curves (an analysis which is beyond the scope of this paper). However, we can assume qualitatively that the SSS durations would be shortened with presumably a slightly later detection on the rise in luminosity (i.e. a longer \tonk) and an earlier \toff because the declining source flux would fall below the detection threshold sooner.

Such behaviour is not consistent with the observations in Fig.\,\ref{fig:ton_toff_bd}. For short turn-on times, the turn-off times of disk novae appear to be on average longer than for bulge novae with comparable \tonk. This discrepancy increases slightly if we assume that the disk turn-on times might be delayed. For longer turn-on times, disk novae seem to turn-off earlier as SSSs, but it is difficult to imagine how absorption alone could act selectively only on slow novae. Overall, we did not find evidence of an obvious observational bias that could explain Fig.\,\ref{fig:ton_toff_bd}.

\subsection{Completeness simulation}
\label{sec:discuss_sim}
In \mzk, we carried out a simulation to determine the intrinsic fraction of \m31 novae with SSS phase based on our monitoring. Here, we repeated this analysis by also including the three monitoring campaigns discussed in this paper (see Table\,\ref{tab:obs}). This means that the simulation was based on 85 individual observations from eight different monitoring seasons (presented in \pek, \pzk, \mbk, and this work). The \xmm ToO observations described in \citet{2012A&A...544A..44H} were not included here because they were not aimed at the \m31 centre.

The details of the simulation are outlined in \mzk. In short, we assumed a theoretically observable WD mass distribution for novae (mainly determined by short recurrence times of intrinsically less frequent high-mass systems) and translated it into an expected SSS turn-on time distribution \citep[following][]{2006ApJS..167...59H}. This was used to estimate an SSS visibility distribution based on Eqn.\,\ref{eqn:ton_toff}. We then took all \m31 novae within our field of view since 1995 (correcting for \xmm source confusion in the innermost part of \m31) and, based on the visibility distribution, we randomly assigned \ton and \toff times to them. A certain fraction $x$ of these novae was then checked against our observations to see if they would have been detected (i.e. their SSS visibility covers at least one observation). By varying this fraction $x$, we adjusted the expected number of detected novae to match the actual number of detections in each campaign using a Markov Chain Monte Carlo procedure. The $x$ that led to the closest agreement with the observed nova counts (using a likelihood criterion) was accepted as the most likely intrinsic fraction of novae with SSS phase. For the resulting Markov chain, we excluded the burn-in data and applied a generous thinning to remove auto-correlation within the chain.

Additionally, we incorporate the asymmetry of the positional distribution of SSS counterparts in this work, as discussed in \mzk. This asymmetry can be seen in Fig.\,\ref{fig:m31_bd}, where most of the novae detected in X-rays are projected onto the far (east) side of \m31. The effect is probably caused by additional foreground absorption by the \m31 disk. Therefore, we excluded optical novae from the near side of the \m31 disk from the simulation (and adjusted the number of actually detected SSSs accordingly). In total, this simulation was based on 234 novae.

\begin{figure}[t!]
  \resizebox{\hsize}{!}{\includegraphics[angle=270]{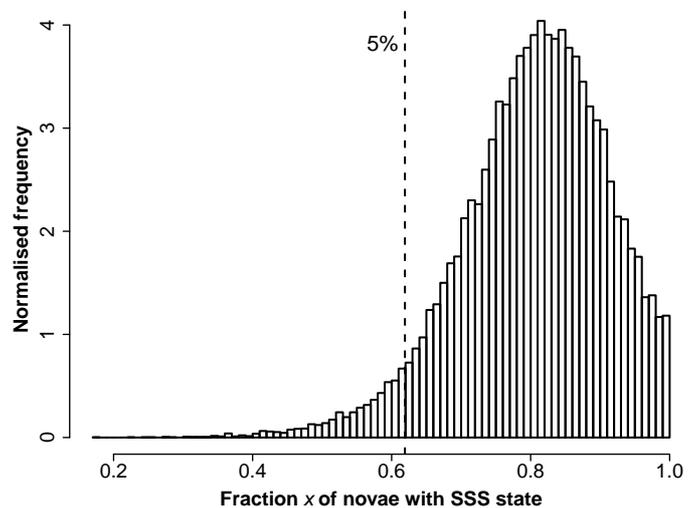}}
  \caption{Simulation: fraction $x$ of novae with intrinsic SSS state.}
  \label{fig:sim}
\end{figure}

The results are shown in Fig.\,\ref{fig:sim}, which plots the distribution of the SSS fraction $x$. The median of this distribution is at 0.81 with the 95\% confidence limit at 0.62, and it can be approximated by a Gaussian truncated at 1 with a standard deviation of 0.1. Given our assumptions, we can therefore say that the intrinsic fraction of novae with SSS state is larger than 62\% with 95\% confidence. The most likely value is at 81\%. Still, we cannot exclude that all novae experience an SSS phase and that the relatively low detection fraction is caused by the inevitably incomplete observational coverage.

%
%

%

A mechanism that could stop \m31 novae from being detected in X-rays is the self-absorption of the SSS flux by the ejected material (i.e. $\eton > \etoff$). The correlation plot in Fig.\,\ref{fig:corr}a provides no strong suggestion of a continuum distribution that may spread into the unobservable region.

\section{Summary}
\label{sec:summary}
%
This paper presents the results of dedicated monitoring observations of the \m31 central area with \xmm and \chandrak. We discovered 17 new X-ray counterparts of optical novae and detected 24 novae in total. Several individual objects were discussed in detail because they either displayed particularly interesting spectral long-term evolutions or showed X-ray light curves that featured noteworthy variability on time scales of hours.

Within the bigger picture, our new results increased the total number of \m31 novae with X-ray detection to 79. A thorough analysis of the extended catalogue yielded several interesting results: A number of correlations between optical and X-ray parameters, as first seen in \m31 data in \mzk, was shown to be stable, and the uncertainties of the corresponding power law models were reduced considerably. We found evidence of multi-parameter relations dominated by a single physical parameter. We suggested interpretations and implications from this behaviour. The well-defined power law relationships allowed us to examine various outliers. We found that there still is evidence of different X-ray characteristics of bulge vs disk novae in \m31 although not all population results from \mz could be confirmed. Last but not least, we cannot exclude that all optical \m31 novae show a SSS state based on a simple simulation, although our observations suggest that 20\% of them might not.

Finally, we wish to emphasise the unique role that \m31 has always played and will hopefully also play in the future for large scale surveys and population studies of novae. The size and proximity of our neighbour galaxy make it an ideal target for studying a nova sample homogeneous in distance but diverse in properties. The coming decades undoubtedly will see large scale optical surveys becoming more popular, and nova discoveries in nearby galaxies will increase in frequency. In the past years, we have seen the fraction of \m31 novae with spectroscopic follow-up increase sharply, and several different groups are now providing optical spectra soon after discovery. It is a golden age for nova studies, and if X-ray observations can keep up with their optical counterparts, then the discovery potential will be high.

\begin{acknowledgements}
We thank the anonymouos referee for comments that helped to improve the paper. We wish to thank A.~W. Shafter and S.~N. Shore for fruitful discussions of preliminary results. We acknowledge the use of Swift information for the Galactic nova V723~Cas, based on monitoring observations obtained by J.-U. Ness and G. Schwarz. The X-ray work is based in part on observations with \xmmk, an ESA Science Mission with instruments and contributions directly funded by ESA Member States and NASA. The \xmm project is supported by the Bundesministerium f\"{u}r Wirtschaft und Technologie / Deutsches Zentrum f\"{u}r Luft- und Raumfahrt (BMWI/DLR FKZ 50 OX 0001) and the Max-Planck Society. M. Henze acknowledges support from the BMWI/DLR, FKZ 50 OR 1010, and from an ESA fellowship. M. Hernanz acknowledges the Spanish MICINN project AYA2011-24704 and FEDER funds. GS acknowledges MICINN grants AYA2010-15685 and AYA2011-23102, Government of Catalonia grant 2009SGR-1002, and the ESF EUROCORES Program EuroGENESIS through the MICINN grant EUI2009-04167.

\end{acknowledgements}

\bibliographystyle{aa}

%
\begin{table*}[!t]
\begin{center}
\caption[]{Observations of the \m31 monitoring.}
\begin{tabular}{lrrrrrrrrr}
\hline\noalign{\smallskip}
\hline\noalign{\smallskip}
\multicolumn{1}{l}{Telescope/Instrument} & \multicolumn{1}{r}{ObsID} & \multicolumn{4}{c}{Exposure Time$^a$ [ks]}
& \multicolumn{1}{c}{Start Date$^b$} & \multicolumn{1}{c}{JD$^b$} & \multicolumn{2}{c}{Offset$^c$}\\
\noalign{\smallskip}
 & & \multicolumn{1}{r}{pn} & \multicolumn{1}{r}{MOS1} & \multicolumn{1}{r}{MOS2} 
& \multicolumn{1}{r}{HRC-I} & \multicolumn{1}{c}{[UT]} & \multicolumn{1}{l}{2\,450\,000+} 
& \multicolumn{1}{r}{RA [$\arcsec$]} & \multicolumn{1}{r}{Dec [$\arcsec$]}\\
\noalign{\smallskip}\hline\noalign{\smallskip}\hline\noalign{\smallskip}
	\multicolumn{2}{c}{\textit{2009/10}} & & & & & & & & \\ \noalign{\smallskip}\hline\noalign{\smallskip}
	\chandra HRC-I & 10882 & & & & 18.8 & 2009-11-07.23 & 5142.73 & -0.3 & 0.4 \\
	\chandra HRC-I & 10883 & & & & 18.3 & 2009-11-16.24 & 5151.74 & -0.4 & 0.3 \\
	\chandra HRC-I & 10884 & & & & 18.4 & 2009-11-27.63 & 5163.13 & -0.3 & 0.3 \\
	\chandra HRC-I & 10885 & & & & 18.3 & 2009-12-08.94 & 5174.44 & -0.4 & 0.2 \\
	\chandra HRC-I & 10886 & & & & 18.3 & 2009-12-17.90 & 5183.40 & -0.4 & 0.1 \\
	\xmm EPIC & 0600660201 & 14.4 & 18.1 & 18.1 & & 2009-12-28.53 & 5194.03 & 1.7 & 0.3 \\
	\xmm EPIC & 0600660301 & 13.2 & 16.7 & 16.8 & & 2010-01-07.32 & 5203.82 & 1.8 & 0.3 \\
	\xmm EPIC & 0600660401 & 10.0 & 16.6 & 16.7 & & 2010-01-15.53 & 5212.03 & -0.3 & 1.5 \\
	\xmm EPIC & 0600660501 & 10.2 & 16.3 & 16.5 & & 2010-01-25.11 & 5221.61 & -0.1 & 1.4 \\
	\xmm EPIC & 0600660601 & 12.0 & 16.6 & 16.5 & & 2010-02-02.11 & 5229.61 & -0.4 & 1.5 \\
	\chandra HRC-I & 11808 & & & & 17.1 & 2010-02-15.86 & 5243.36 & -0.4 & 0.0 \\
	\chandra HRC-I & 11809 & & & & 18.4 & 2010-02-26.27 & 5253.77 & -0.4 & 0.0 \\
	\noalign{\smallskip}\hline\noalign{\smallskip}
	\multicolumn{2}{c}{\textit{2010/11}} & & & & & & & & \\ \noalign{\smallskip}\hline\noalign{\smallskip}
	\chandra HRC-I & 12110 & & & & 20.0 & 2010-11-14.17 & 5514.67 & -0.8 & 0.2 \\
	\chandra HRC-I & 12111 & & & & 19.9 & 2010-11-23.18 & 5523.68 & -0.7 & 0.3 \\
	\chandra HRC-I & 12112 & & & & 19.9 & 2010-12-03.66 & 5534.16 & -0.6 & 0.2 \\
	\chandra HRC-I & 12113 & & & & 19.0 & 2010-12-12.56 & 5543.06 & -0.5 & 0.1 \\
	\chandra HRC-I & 12114 & & & & 20.0 & 2010-12-22.18 & 5552.68 & -0.5 & 0.2 \\
	\xmm EPIC & 0650560201 & 18.7 & 23.5 & 24.1 & & 2010-12-26.43 & 5556.93 & 0.9 & 0.3 \\
	\xmm EPIC & 0650560301 & 21.6 & 30.6 & 31.4 & & 2011-01-04.76 & 5566.26 & -0.2 & 0.3 \\
	\xmm EPIC & 0650560401 & 11.9 & 16.3 & 18.6 & & 2011-01-15.01 & 5576.51 & 1.0 & 1.0 \\
	\xmm EPIC & 0650560501 & 6.2 & 20.7 & 21.7 & & 2011-01-25.30 & 5586.80 & 0.1 & 0.8 \\
	\xmm EPIC & 0650560601 & 16.2 & 22.9 & 23.0 & & 2011-02-04.00 & 5596.50 & -0.2 & 0.6 \\
	\chandra HRC-I & 13178 & & & & 17.5 & 2011-02-17.15 & 5609.65 & -0.4 & 0.0 \\
	\chandra HRC-I & 13179 & & & & 17.5 & 2011-02-27.25 & 5619.75 & -0.2 & -0.1 \\
	\chandra HRC-I & 13180 & & & & 17.3 & 2011-03-10.12 & 5630.62 & 0.0 & -0.2 \\
	\noalign{\smallskip}\hline\noalign{\smallskip}
	\multicolumn{2}{c}{\textit{2011/12}} & & & & & & & & \\ \noalign{\smallskip}\hline\noalign{\smallskip}
	\chandra HRC-I & 13227 & & & & 20.0 & 2011-11-12.10 & 5877.60 & -0.4 & 0.1 \\
	\chandra HRC-I & 13228 & & & & 19.0 & 2011-11-21.22 & 5886.72 & -0.5 & 0.1 \\
	\chandra HRC-I & 13229 & & & & 19.6 & 2011-11-30.98 & 5896.48 & -0.2 & 0.4 \\
	\chandra HRC-I & 13230 & & & & 18.9 & 2011-12-11.56 & 5907.06 & -0.3 & 0.4 \\
	\chandra HRC-I & 13231 & & & & 19.5 & 2011-12-20.33 & 5915.83 & -0.4 & -0.2 \\
	\xmm EPIC & 0674210201 & 16.7 & 20.3 & 20.3 & & 2011-12-28.05 & 5923.55 & -1.7 & 0.1 \\
	\xmm EPIC & 0674210301 & 13.6 & 16.8 & 16.8 & & 2012-01-07.12 & 5933.62 & 0.4 & 0.2 \\
	\xmm EPIC & 0674210401 & 15.9 & 19.4 & 19.4 & & 2012-01-15.62 & 5942.12 & 0.3 & 0.7 \\
	\xmm EPIC & 0674210501 & 13.6 & 16.8 & 16.8 & &  2012-01-21.51 & 5948.01 & -0.1 & 1.5 \\
	\xmm EPIC & 0674210601 & 17.8 & 20.9 & 21.2 & & 2012-01-31.10 & 5957.60 & -1.1 & 0.1 \\
	\chandra HRC-I & 13278 & & & & 19.0 & 2012-02-17.76 & 5973.26 & -0.5 & -0.2 \\
	\chandra HRC-I & 13279 & & & & 18.8 & 2012-02-28.26 & 5983.76 & -0.2 & -0.3 \\
	\chandra HRC-I & 13280 & & & & 19.3 & 2012-03-13.21 & 5999.71 & -0.1 & -0.5 \\
	\chandra HRC-I & 13281 & & & & 18.9 & 2012-06-01.90 & 6080.40 & 0.4 & 0.0 \\
\noalign{\smallskip} \hline
\end{tabular}
\label{tab:obs}
\end{center}
Notes:\hspace{0.3cm} $^a $: Dead-time corrected; for \xmm EPIC after screening for high background.\\
\hspace*{1.1cm} $^b $: Start time of observations; for \xmm EPIC the pn start time was used.\\
\hspace*{1.1cm} $^c $: Offset of image WCS (world coordinate system) to the WCS of the catalogue by \citet{2002ApJ...578..114K}.\\
\end{table*}
%

%
\begin{table*}[t!]
\begin{center}
\caption[]{\xmm and \chandra measurements of \m31 optical nova candidates known from \mz and still detected here.}
\begin{tabular}{lrrlrrrl}
\hline\noalign{\smallskip}
\hline\noalign{\smallskip}
\multicolumn{3}{l}{Optical nova candidate} & \multicolumn{3}{l}{X-ray measurements} \\
\noalign{\smallskip}\hline\noalign{\smallskip}
\multicolumn{1}{l}{Name} & \multicolumn{1}{c}{RA~~~(h:m:s)$^a$} & \multicolumn{1}{c}{MJD$^b$} & \multicolumn{1}{c}{$D^c$} 
& \multicolumn{1}{c}{Observation$^d$} & \multicolumn{1}{c}{$\Delta t^e$} & \multicolumn{1}{c}{$L_{\rm 50}^f$}
& \multicolumn{1}{l}{Comment$^g$} \\
M31N & \multicolumn{1}{c}{Dec~(d:m:s)$^a$} & \multicolumn{1}{c}{(d)} & (\arcsec)  & \multicolumn{1}{c}{ID} &\multicolumn{1}{c}{(d)} &\multicolumn{1}{c}{(10$^{36}$ erg s$^{-1}$)} & \\ 
\noalign{\smallskip}\hline\noalign{\smallskip}
 1996-08b& 00:42:55.20 & 50307.0 & 1.7 &mrg3 (HRC-I) & 4835.2 & $2.2\pm0.5$ & \\
        & +41:20:46.0 &      &           &mrg3 (EPIC) & 4886.5 & $0.9\pm0.2$ & \\
        &            &      & & mrg4 (HRC-I) & 5207.2 & $< 11.7$ & \\
        &            &      & & mrg4 (EPIC) & 5249.4 & $< 0.3$ & \\
\noalign{\smallskip}
 2001-10a& 00:43:03.31 & 52185.91 & 0.1 &10882 (HRC-I) & 2956.3 & $11.1\pm2.1$ & \\
        & +41:12:11.5 &      & 0.3 &10883 (HRC-I) & 2965.3 & $5.3\pm1.6$ & \\
        &      &      & 1.6 &10884 (HRC-I) & 2976.7 & $5.4\pm1.6$ & \\
        &      &      & 2.3 &10885 (HRC-I) & 2988.0 & $3.1\pm1.3$ & \\
        &      &      & 1.5 &10886 (HRC-I) & 2997.0 & $8.8\pm2.0$ & \\
        &      &      & 1.4 &0600660201 (EPIC) & 3007.6 & $2.8\pm0.6$ & \\
        &      &      &           &0600660301 (EPIC) & 3017.4 & $2.8\pm1.0$ & \\
        &      &      &           &0600660401 (EPIC) & 3025.6 & $3.0\pm0.7$ & \\
        &      &      &           &0600660501 (EPIC) & 3035.2 & $3.3\pm0.7$ & \\
        &      &      &           &0600660601 (EPIC) & 3043.2 & $1.9\pm0.5$ & \\
        &      &      & 0.7 &11808 (HRC-I) & 3056.9 & $3.6\pm1.3$ & \\
        &      &      &           &11809 (HRC-I) & 3067.4 & $< 15.8$ & \\
        &      &      & 1.9 &12110 (HRC-I) & 3328.3 & $6.5\pm1.6$ & \\
        &      &      &           &12111 (HRC-I) & 3337.3 & $< 15.7$ & \\
        &      &      & 1.0 &12112 (HRC-I) & 3347.8 & $5.7\pm1.5$ & \\
        &      &      & 0.7 &12113 (HRC-I) & 3356.6 & $4.6\pm1.4$ & \\
        &      &      & 1.8 &12114 (HRC-I) & 3366.3 & $2.8\pm1.2$ & \\
        &      &      & 1.9 &0650560201 (EPIC) & 3370.5 & $0.1\pm0.0$ & \\
        &      &      &           &0650560301 (EPIC) & 3379.8 & $2.0\pm0.8$ & \\
        &      &      &           &0650560401 (EPIC) & 3390.1 & $1.5\pm0.5$ & \\
        &      &      &           &0650560501 (EPIC) & 3400.4 & $< 2.8$ & \\
        &      &      &           &0650560601 (EPIC) & 3410.1 & $1.2\pm0.4$ & \\
        &      &      &           &13178 (HRC-I) & 3423.2 & $< 7.8$ & \\
        &      &      & 2.2 &13179 (HRC-I) & 3433.3 & $4.3\pm1.4$ & \\
        &      &      &           &13180 (HRC-I) & 3444.2 & $< 12.6$ & \\
        &      &      & & 0655620301 (EPIC) & 3588.3 & $< 0.7$ & ToO\\
\noalign{\smallskip}
 2003-08c& 00:42:41.20 & 52878.0 & 0.2 &mrg3 (HRC-I) & 2264.2 & $0.7\pm0.2$ & \\
        & +41:16:16.0 &      & 0.1 &mrg4 (HRC-I) & 2636.2 & $1.1\pm0.2$ & \\
        &            &      & 0.4 &mrg5 (HRC-I) & 2999.1 & $0.2\pm0.1$ & \\
\noalign{\smallskip}
 2004-01b& 00:42:41.18 & 53005.74 & 0.2 &10882 (HRC-I) & 2136.5 & $11.4\pm1.6$ & \\
        & +41:15:45.0 &      & 0.1 &10883 (HRC-I) & 2145.5 & $9.7\pm1.5$ & \\
        &      &      & 0.2 &10884 (HRC-I) & 2156.9 & $6.0\pm1.2$ & \\
        &      &      & 0.2 &10885 (HRC-I) & 2168.2 & $9.2\pm1.4$ & \\
        &      &      & 0.1 &10886 (HRC-I) & 2177.2 & $11.9\pm1.6$ & \\
        &      &      & 0.2 &11808 (HRC-I) & 2237.1 & $7.8\pm1.4$ & \\
        &      &      & 0.3 &11809 (HRC-I) & 2247.5 & $3.9\pm0.9$ & \\
        &      &      &           &12110 (HRC-I) & 2508.4 & $< 1.5$ & \\
        &      &      & 0.2 &12111 (HRC-I) & 2517.4 & $1.5\pm0.6$ & \\
        &      &      &           &12112 (HRC-I) & 2527.9 & $< 4.2$ & \\
        &      &      &           &12113 (HRC-I) & 2536.8 & $< 3.7$ & \\
        &      &      &           &12114 (HRC-I) & 2546.4 & $< 2.7$ & \\
        &      &      & 0.4 &13178 (HRC-I) & 2603.4 & $1.5\pm0.6$ & \\
        &      &      &           &13179 (HRC-I) & 2613.5 & $< 3.1$ & \\
        &      &      & 0.3 &13180 (HRC-I) & 2624.4 & $1.9\pm0.7$ & \\
        &      &      & 0.2 &13227 (HRC-I) & 2871.4 & $5.6\pm1.1$ & \\
        &      &      & 0.1 &13228 (HRC-I) & 2880.5 & $6.8\pm1.2$ & \\
        &      &      & 0.2 &13229 (HRC-I) & 2890.2 & $7.2\pm1.2$ & \\
        &      &      & 0.1 &13230 (HRC-I) & 2900.8 & $5.6\pm1.1$ & \\
        &      &      & 0.2 &13231 (HRC-I) & 2909.6 & $6.2\pm1.1$ & \\
        &      &      & 0.1 &13278 (HRC-I) & 2967.0 & $7.0\pm1.2$ & \\
        &      &      & 0.1 &13279 (HRC-I) & 2977.5 & $5.4\pm1.1$ & \\
        &      &      & 0.2 &13280 (HRC-I) & 2993.5 & $6.9\pm1.2$ & \\
        &      &      & 0.1 &13281 (HRC-I) & 3074.2 & $4.8\pm1.1$ & \\
\noalign{\smallskip}
\hline
\noalign{\smallskip}
\end{tabular}
\label{tab:novae_old_lum}
\end{center}
\end{table*}
%

\begin{table*}[t]
\begin{center}
\addtocounter{table}{-1}
\caption[]{continued.}
\begin{tabular}{lrrlrrrl}
\hline\noalign{\smallskip}
\hline\noalign{\smallskip}
\multicolumn{3}{l}{Optical nova candidate} & \multicolumn{3}{l}{X-ray measurements} \\
\noalign{\smallskip}\hline\noalign{\smallskip}
\multicolumn{1}{l}{Name} & \multicolumn{1}{c}{RA~~~(h:m:s)$^a$} & \multicolumn{1}{c}{MJD$^b$} & \multicolumn{1}{c}{$D^c$} 
& \multicolumn{1}{c}{Observation$^d$} & \multicolumn{1}{c}{$\Delta t^e$} & \multicolumn{1}{c}{$L_{\rm 50}^f$}
& \multicolumn{1}{l}{Comment$^g$} \\
M31N & \multicolumn{1}{c}{Dec~(d:m:s)$^a$} & \multicolumn{1}{c}{(d)} & (\arcsec)  & \multicolumn{1}{c}{ID} &\multicolumn{1}{c}{(d)} &\multicolumn{1}{c}{(10$^{36}$ erg s$^{-1}$)} & \\ 
\noalign{\smallskip}\hline\noalign{\smallskip}
 2004-05b& 00:42:37.04 & 53143.06 & 0.3 &10882 (HRC-I) & 1999.2 & $9.1\pm1.4$ & \\
        & +41:14:28.5 &      & 0.4 &10883 (HRC-I) & 2008.2 & $10.7\pm1.6$ & \\
        &      &      & 0.4 &10884 (HRC-I) & 2019.6 & $7.4\pm1.3$ & \\
        &      &      & 0.5 &10885 (HRC-I) & 2030.9 & $5.3\pm1.1$ & \\
        &      &      & 0.4 &10886 (HRC-I) & 2039.8 & $5.1\pm1.1$ & \\
        &      &      &           &0600660201 (EPIC) & 2050.5 & $3.6\pm0.7$ & \\
        &      &      &           &0600660301 (EPIC) & 2060.3 & $2.5\pm0.7$ & \\
        &      &      &           &0600660401 (EPIC) & 2068.5 & $< 2.4$ & \\
        &      &      &           &0600660501 (EPIC) & 2078.0 & $2.2\pm0.7$ & \\
        &      &      &           &0600660601 (EPIC) & 2086.0 & $2.0\pm0.7$ & \\
        &      &      & 0.3 &11808 (HRC-I) & 2099.8 & $1.7\pm0.7$ & \\
        &      &      &           &11809 (HRC-I) & 2110.2 & $< 5.5$ & \\
        &      &      & & mrg4 (HRC-I) & 2371.1 & $< 1.3$ & \\
        &      &      & & mrg4 (EPIC) & 2413.4 & $< 0.1$ & \\
\noalign{\smallskip}
 2006-06b& 00:42:32.77 & 53867.07 & 0.3 &10882 (HRC-I) & 1275.2 & $6.7\pm1.2$ & \\
        & +41:16:49.1 &      & 0.4 &10883 (HRC-I) & 1284.2 & $5.5\pm1.1$ & \\
        &      &      & 0.3 &10884 (HRC-I) & 1295.6 & $9.8\pm1.5$ & \\
        &      &      & 0.3 &10885 (HRC-I) & 1306.9 & $6.9\pm1.3$ & \\
        &      &      & 0.3 &10886 (HRC-I) & 1315.8 & $5.2\pm1.1$ & \\
        &      &      & 0.8 &0600660201 (EPIC) & 1326.5 & $3.3\pm0.6$ & \\
        &      &      &           &0600660301 (EPIC) & 1336.2 & $4.4\pm0.8$ & \\
        &      &      &           &0600660401 (EPIC) & 1344.5 & $4.6\pm0.9$ & \\
        &      &      & 2.8 &0600660501 (EPIC) & 1354.0 & $5.9\pm0.8$ & \\
        &      &      &           &0600660601 (EPIC) & 1362.0 & $4.5\pm0.9$ & \\
        &      &      & 0.2 &11808 (HRC-I) & 1375.8 & $8.3\pm1.4$ & \\
        &      &      & 0.2 &11809 (HRC-I) & 1386.2 & $8.4\pm1.4$ & \\
        &      &      & 0.2 &12110 (HRC-I) & 1647.1 & $5.8\pm1.1$ & \\
        &      &      & 0.1 &12111 (HRC-I) & 1656.1 & $6.7\pm1.2$ & \\
        &      &      & 0.5 &12112 (HRC-I) & 1666.6 & $6.0\pm1.2$ & \\
        &      &      & 0.3 &12113 (HRC-I) & 1675.5 & $4.9\pm1.1$ & \\
        &      &      & 0.3 &12114 (HRC-I) & 1685.1 & $7.3\pm1.3$ & \\
        &      &      & 1.1 &0650560201 (EPIC) & 1689.4 & $0.1\pm0.0$ & \\
        &      &      & 0.5 &0650560301 (EPIC) & 1698.7 & $4.2\pm0.5$ & \\
        &      &      & 1.1 &0650560401 (EPIC) & 1708.9 & $4.1\pm0.7$ & \\
        &      &      &           &0650560501 (EPIC) & 1719.2 & $5.1\pm1.2$ & \\
        &      &      & 0.8 &0650560601 (EPIC) & 1728.9 & $4.1\pm0.6$ & \\
        &      &      & 0.2 &13178 (HRC-I) & 1742.1 & $6.7\pm1.3$ & \\
        &      &      & 0.3 &13179 (HRC-I) & 1752.2 & $3.3\pm0.9$ & \\
        &      &      & 0.2 &13180 (HRC-I) & 1763.1 & $4.0\pm1.0$ & \\
        &      &      & 0.2 &13227 (HRC-I) & 2010.0 & $1.9\pm0.7$ & \\
        &      &      & 0.5 &13228 (HRC-I) & 2019.2 & $2.6\pm0.8$ & \\
        &      &      & 0.2 &13229 (HRC-I) & 2028.9 & $1.9\pm0.7$ & \\
        &      &      & 0.2 &13230 (HRC-I) & 2039.5 & $2.4\pm0.8$ & \\
        &      &      & 0.2 &13231 (HRC-I) & 2048.3 & $1.8\pm0.6$ & \\
        &      &      &           &0674210201 (EPIC) & 2056.0 & $< 2.3$ & \\
        &      &      &           &0674210301 (EPIC) & 2066.1 & $< 2.9$ & \\
        &      &      &           &0674210401 (EPIC) & 2074.6 & $0.7\pm0.4$ & \\
        &      &      &           &0674210501 (EPIC) & 2080.4 & $< 2.5$ & \\
        &      &      &           &0674210601 (EPIC) & 2090.0 & $< 2.2$ & \\
        &      &      &           &13278 (HRC-I) & 2105.7 & $< 3.2$ & \\
        &      &      &           &13279 (HRC-I) & 2116.2 & $< 1.6$ & \\
        &      &      &           &13280 (HRC-I) & 2132.1 & $< 4.7$ & \\
        &      &      &           &13281 (HRC-I) & 2212.8 & $< 1.4$ & \\
\noalign{\smallskip}
\hline
\noalign{\smallskip}
\end{tabular}
\end{center}
\end{table*}
%

\begin{table*}[t]
\begin{center}
\addtocounter{table}{-1}
\caption[]{continued.}
\begin{tabular}{lrrlrrrl}
\hline\noalign{\smallskip}
\hline\noalign{\smallskip}
\multicolumn{3}{l}{Optical nova candidate} & \multicolumn{3}{l}{X-ray measurements} \\
\noalign{\smallskip}\hline\noalign{\smallskip}
\multicolumn{1}{l}{Name} & \multicolumn{1}{c}{RA~~~(h:m:s)$^a$} & \multicolumn{1}{c}{MJD$^b$} & \multicolumn{1}{c}{$D^c$} 
& \multicolumn{1}{c}{Observation$^d$} & \multicolumn{1}{c}{$\Delta t^e$} & \multicolumn{1}{c}{$L_{\rm 50}^f$}
& \multicolumn{1}{l}{Comment$^g$} \\
M31N & \multicolumn{1}{c}{Dec~(d:m:s)$^a$} & \multicolumn{1}{c}{(d)} & (\arcsec)  & \multicolumn{1}{c}{ID} &\multicolumn{1}{c}{(d)} &\multicolumn{1}{c}{(10$^{36}$ erg s$^{-1}$)} & \\ 
\noalign{\smallskip}\hline\noalign{\smallskip}
 2007-02b& 00:41:40.32 & 54134.8 & 3.0 &0600660201 (EPIC) & 1058.7 & $13.6\pm5.0$ & \\
        & +41:14:33.5 &      & 2.2 &0600660301 (EPIC) & 1068.5 & $7.9\pm3.2$ & \\
        &      &      & 2.2 &0600660501 (EPIC) & 1086.3 & $8.5\pm1.7$ & \\
        &      &      & 0.5 &0600660601 (EPIC) & 1094.3 & $6.4\pm1.0$ & \\
        &      &      & & mrg4 (EPIC) & 1421.6 & $< 0.7$ & \\
        &      & &  & & & \\
\noalign{\smallskip}
\hline
\noalign{\smallskip}
\end{tabular}
\end{center}
Notes: \hspace{0.2cm} $^a$: RA, Dec are given in J2000.0; $^b $: Modified Julian Date of optical outburst; MJD = JD - 2\,400\,000.5; $^c$: Distance in arcsec between optical and X-ray source; $^d$: mrg3/4/5 (HRC-I/EPIC) indicates merged data of all HRC-I/EPIC observations during 2009/10, 2010/11 or 2011/12; $^e $: Time after observed start of optical outburst; $^f $: unabsorbed equivalent luminosity in the 0.2--10.0 keV band assuming a 50 eV black body spectrum with Galactic foreground absorption (luminosity errors are 1$\sigma$, upper limits are 3$\sigma$); $^g $: SSS or SSS-HR indicate X-ray sources classified as supersoft based on \xmm spectra or \chandra hardness ratios, respectively. The comment ToO refers to the observations presented in \citet{2012A&A...544A..44H}.\\
\end{table*}
%

\begin{table*}[t!]
\begin{center}
\caption[]{\m31 optical novae with \xmm and \chandra counterparts discovered in this work.}
\begin{tabular}{lrrlrrrl}
\hline\noalign{\smallskip}
\hline\noalign{\smallskip}
\multicolumn{3}{l}{Optical nova candidate} & \multicolumn{3}{l}{X-ray measurements} \\
\noalign{\smallskip}\hline\noalign{\smallskip}
\multicolumn{1}{l}{Name} & \multicolumn{1}{c}{RA~~~(h:m:s)$^a$} & \multicolumn{1}{c}{MJD$^b$} & \multicolumn{1}{c}{$D^c$} 
& \multicolumn{1}{c}{Observation$^d$} & \multicolumn{1}{c}{$\Delta t^e$} & \multicolumn{1}{c}{$L_{\rm 50}^f$}
& \multicolumn{1}{l}{Comment$^g$} \\
M31N & \multicolumn{1}{c}{Dec~(d:m:s)$^a$} & \multicolumn{1}{c}{(d)} & (\arcsec) & \multicolumn{1}{c}{ID} &\multicolumn{1}{c}{(d)} &\multicolumn{1}{c}{(10$^{36}$ erg s$^{-1}$)} & \\ 
\noalign{\smallskip}\hline\noalign{\smallskip}
 2002-08b& 00:42:48.66 & 52512.06 & &mrg1 (HRC-I) & 1899.6 & $< 0.4$ & \\
        & +41:16:26.3 & &    &mrg2 (HRC-I) & 2266.3 & $< 2.2$ & \\
        &      &      & 0.3 &mrg3 (HRC-I) & 2630.2 & $0.4\pm0.1$ & \\
        &      &      & 0.6 &mrg4 (HRC-I) & 3002.1 & $1.2\pm0.2$ & \\
        &      &      & 0.4 &mrg5 (HRC-I) & 3365.0 & $0.8\pm0.2$ & \\
\end{tabular}
\label{tab:novae_new_lum}
\end{center}
\end{table*}
%

\begin{table*}[t]
\begin{center}
\addtocounter{table}{-1}
\caption[]{continued.}
\begin{tabular}{lrrlrrrl}
\hline\noalign{\smallskip}
\hline\noalign{\smallskip}
\multicolumn{3}{l}{Optical nova candidate} & \multicolumn{3}{l}{X-ray measurements} \\
\noalign{\smallskip}\hline\noalign{\smallskip}
\multicolumn{1}{l}{Name} & \multicolumn{1}{c}{RA~~~(h:m:s)$^a$} & \multicolumn{1}{c}{MJD$^b$} & \multicolumn{1}{c}{$D^c$} 
& \multicolumn{1}{c}{Observation$^d$} & \multicolumn{1}{c}{$\Delta t^e$} & \multicolumn{1}{c}{$L_{\rm 50}^f$}
& \multicolumn{1}{l}{Comment$^g$} \\
M31N & \multicolumn{1}{c}{Dec~(d:m:s)$^a$} & \multicolumn{1}{c}{(d)} & (\arcsec) & \multicolumn{1}{c}{ID} &\multicolumn{1}{c}{(d)} &\multicolumn{1}{c}{(10$^{36}$ erg s$^{-1}$)} & \\ 
\noalign{\smallskip}\hline\noalign{\smallskip}
 2009-08e& 00:42:36.16 & 55068.9 &           &10882 (HRC-I) & 73.3 & $< 1.3$ & \\
        & +41:18:02.4 &      &           &10883 (HRC-I) & 82.3 & $< 1.3$ & \\
        &      &      &           &10884 (HRC-I) & 93.7 & $< 3.3$ & \\
        &      &      &           &10885 (HRC-I) & 105.0 & $< 1.9$ & \\
        &      &      &           &10886 (HRC-I) & 114.0 & $< 2.8$ & \\
\end{tabular}
\end{center}
\end{table*}

\begin{table*}[t]
\begin{center}
\addtocounter{table}{-1}
\caption[]{continued.}
\begin{tabular}{lrrlrrrl}
\hline\noalign{\smallskip}
\hline\noalign{\smallskip}
\multicolumn{3}{l}{Optical nova candidate} & \multicolumn{3}{l}{X-ray measurements} \\
\noalign{\smallskip}\hline\noalign{\smallskip}
\multicolumn{1}{l}{Name} & \multicolumn{1}{c}{RA~~~(h:m:s)$^a$} & \multicolumn{1}{c}{MJD$^b$} & \multicolumn{1}{c}{$D^c$} 
& \multicolumn{1}{c}{Observation$^d$} & \multicolumn{1}{c}{$\Delta t^e$} & \multicolumn{1}{c}{$L_{\rm 50}^f$}
& \multicolumn{1}{l}{Comment$^g$} \\
M31N & \multicolumn{1}{c}{Dec~(d:m:s)$^a$} & \multicolumn{1}{c}{(d)} & (\arcsec) & \multicolumn{1}{c}{ID} &\multicolumn{1}{c}{(d)} &\multicolumn{1}{c}{(10$^{36}$ erg s$^{-1}$)} & \\ 
\noalign{\smallskip}\hline\noalign{\smallskip}
 2010-10f& 00:42:43.58 & 55481.19 & 1.5 &12110 (HRC-I) & 33.0 & $79.0\pm4.2$ & GC\\
        & +41:12:42.6 &      & 2.3 &12111 (HRC-I) & 42.0 & $1.8\pm0.7$ & \\
        &      &      & 1.6 &12112 (HRC-I) & 52.5 & $2.9\pm0.9$ & \\
        &      &      &           &12113 (HRC-I) & 61.4 & $< 2.0$ & \\
        &      &      &           &12114 (HRC-I) & 71.0 & $< 4.9$ & \\
\end{tabular}
\end{center}
\end{table*}
%

\begin{table*}[t]
\begin{center}
\addtocounter{table}{-1}
\caption[]{continued.}
\begin{tabular}{lrrlrrrl}
\hline\noalign{\smallskip}
\hline\noalign{\smallskip}
\multicolumn{3}{l}{Optical nova candidate} & \multicolumn{3}{l}{X-ray measurements} \\
\noalign{\smallskip}\hline\noalign{\smallskip}
\multicolumn{1}{l}{Name} & \multicolumn{1}{c}{RA~~~(h:m:s)$^a$} & \multicolumn{1}{c}{MJD$^b$} & \multicolumn{1}{c}{$D^c$} 
& \multicolumn{1}{c}{Observation$^d$} & \multicolumn{1}{c}{$\Delta t^e$} & \multicolumn{1}{c}{$L_{\rm 50}^f$}
& \multicolumn{1}{l}{Comment$^g$} \\
M31N & \multicolumn{1}{c}{Dec~(d:m:s)$^a$} & \multicolumn{1}{c}{(d)} & (\arcsec) & \multicolumn{1}{c}{ID} &\multicolumn{1}{c}{(d)} &\multicolumn{1}{c}{(10$^{36}$ erg s$^{-1}$)} & \\ 
\noalign{\smallskip}\hline\noalign{\smallskip}
 2011-11e& 00:42:38.31 & 55884.7 &           &13228 (HRC-I) & 1.5 & $< 1.1$ & \\
        & +41:16:31.3 &      &           &13229 (HRC-I) & 11.3 & $< 0.7$ & \\
        &      &      &           &13230 (HRC-I) & 21.9 & $< 3.7$ & \\
        &      &      &           &13231 (HRC-I) & 30.6 & $< 1.5$ & \\
        &      &      &           &0674210201 (EPIC) & 38.3 & $< 1.4$ & \\
        &      &      &           &0674210301 (EPIC) & 48.4 & $< 7.8$ & \\
        &      &      &           &0674210401 (EPIC) & 56.9 & $< 8.8$ & \\
        &      &      &           &0674210501 (EPIC) & 62.8 & $< 1.6$ & \\
        &      &      &           &0674210601 (EPIC) & 72.4 & $< 9.9$ & \\
        &      &      & 0.5 &13278 (HRC-I) & 88.1 & $24.1\pm2.3$ & \\
        &      &      & 0.5 &13279 (HRC-I) & 98.6 & $24.4\pm2.3$ & \\
        &      &      & 0.5 &13280 (HRC-I) & 114.5 & $28.9\pm2.5$ & \\
        &      &      &           &13281 (HRC-I) & 195.2 & $< 0.7$ & \\
\noalign{\smallskip}
\hline
\noalign{\smallskip}
\end{tabular}
\end{center}
Notes: \hspace{0.2cm} As for Table\,\ref{tab:novae_old_lum}. The entries mrg1 and mrg2 refer to the merged data of the 2007/8 and 2008/9 campaigns presented in \mz (same abbreviations there). ObsID 13100 indicates the HRC-I observations 13178, 13179 and 13180 merged. The comment GC refers to the nova in the globular cluster Bol~126 \citep{2013A&A...549A.120H}.\\
\end{table*}
%

\begin{table*}[t!]
\begin{center}
\caption[]{Upper limits for non-detected \m31 CNe from \mzk.}
\begin{tabular}{lrrlrrrl}
\hline\noalign{\smallskip}
\hline\noalign{\smallskip}
\multicolumn{3}{l}{Optical nova candidate} & \multicolumn{3}{l}{X-ray measurements} \\
\noalign{\smallskip}\hline\noalign{\smallskip}
\multicolumn{1}{l}{Name} & \multicolumn{1}{c}{RA~~~(h:m:s)$^a$} & \multicolumn{1}{c}{MJD$^b$} & \multicolumn{1}{c}{Observation$^d$} & \multicolumn{1}{c}{$\Delta t^e$} & \multicolumn{1}{c}{$L_{\rm 50}^f$}
& \multicolumn{1}{l}{Comment} \\
M31N & \multicolumn{1}{c}{Dec~(d:m:s)$^a$} & \multicolumn{1}{c}{(d)} & \multicolumn{1}{c}{ID} &\multicolumn{1}{c}{(d)} &\multicolumn{1}{c}{(10$^{36}$ erg s$^{-1}$)} & \\ 
\noalign{\smallskip}\hline\noalign{\smallskip}
 1997-11a& 00:42:42.13 & 50753.05 & mrg3 (HRC-I) & 4389.2 & $< 1.0$ & \\
        & +41:15:10.4 & &  & & & \\
\noalign{\smallskip}
 2008-05a& 00:42:56.84 & 54600.79 & mrg3 (HRC-I) & 541.4 & $< 10.1$ & \\
        & +41:11:52.4 &      & mrg3 (EPIC) & 592.7 & $< 0.1$ & \\
\noalign{\smallskip}
 2008-06a& 00:42:37.72 & 54631.46 & mrg3 (HRC-I) & 510.8 & $< 1.8$ & \\
        & +41:12:30.0 &      & mrg3 (EPIC) & 562.1 & $< 0.1$ & \\
\noalign{\smallskip}
\hline
\noalign{\smallskip}
\end{tabular}
\label{tab:novae_old_non}
\end{center}
Notes: \hspace{0.2cm} As for Table\,\ref{tab:novae_old_lum}.\\
\end{table*}

\clearpage

\begin{table*}[t!]
\begin{center}
\caption[]{Upper limits for \m31 CNe with outburst from about one year before the start of the 2009/10 monitoring until its end.}
\begin{tabular}{lrrrrrl}
\hline\noalign{\smallskip}
\hline\noalign{\smallskip}
\multicolumn{3}{l}{Optical nova candidate} & \multicolumn{4}{l}{X-ray measurements} \\
\noalign{\smallskip}\hline\noalign{\smallskip}
\multicolumn{1}{l}{Name} & \multicolumn{1}{c}{RA~~~(h:m:s)$^a$} & \multicolumn{1}{c}{MJD$^b$}
& \multicolumn{1}{c}{Observation$^d$} & \multicolumn{1}{c}{$\Delta t^e$} & \multicolumn{1}{c}{$L_{\rm 50}^f$}
& \multicolumn{1}{l}{Comment} \\
M31N & \multicolumn{1}{c}{Dec~(d:m:s)$^a$} & \multicolumn{1}{c}{(d)} & \multicolumn{1}{c}{ID} &\multicolumn{1}{c}{(d)} &\multicolumn{1}{c}{(10$^{36}$ erg s$^{-1}$)} & \\ 
\noalign{\smallskip}\hline\noalign{\smallskip}
 2008-10b& 00:43:02.42 & 54745.11 & mrg3 (HRC-I) & 397.1 & $< 1.7$ & \\
        & +41:14:09.9 &      & mrg3 (EPIC) & 448.4 & $< 0.4$ & \\
\noalign{\smallskip}
 2008-10c& 00:42:48.50 & 54759.02 & mrg3 (HRC-I) & 383.2 & $< 1.2$ & \\
        & +41:13:49.8 &      & mrg3 (EPIC) & 434.5 & $< 0.4$ & \\
\noalign{\smallskip}
 2008-11d& 00:42:57.30 & 54794.88 & mrg3 (HRC-I) & 347.4 & $< 1.3$ & \\
\end{tabular}
\end{center}
Notes: \hspace{0.2cm} As for Table\,\ref{tab:novae_old_lum}.\\
\vspace{1cm}
\normalsize
\end{table*}
%

\begin{table*}[t!]
\begin{center}
\caption[]{Upper limits for \m31 CNe with outburst from about one year before the start of the 2010/11 monitoring until its end.}
\begin{tabular}{lrrrrrl}
\hline\noalign{\smallskip}
\hline\noalign{\smallskip}
\multicolumn{3}{l}{Optical nova candidate} & \multicolumn{4}{l}{X-ray measurements} \\
\noalign{\smallskip}\hline\noalign{\smallskip}
\multicolumn{1}{l}{Name} & \multicolumn{1}{c}{RA~~~(h:m:s)$^a$} & \multicolumn{1}{c}{MJD$^b$}
& \multicolumn{1}{c}{Observation$^d$} & \multicolumn{1}{c}{$\Delta t^e$} & \multicolumn{1}{c}{$L_{\rm 50}^f$}
& \multicolumn{1}{l}{Comment} \\
M31N & \multicolumn{1}{c}{Dec~(d:m:s)$^a$} & \multicolumn{1}{c}{(d)} & \multicolumn{1}{c}{ID} &\multicolumn{1}{c}{(d)} &\multicolumn{1}{c}{(10$^{36}$ erg s$^{-1}$)} & \\ 
\noalign{\smallskip}\hline\noalign{\smallskip}
 2009-10c& 00:42:45.76 & 55113.08 & mrg4 (HRC-I) & 401.1 & $< 1.3$ & \\
        & +41:15:57.1 &      & mrg4 (EPIC) & 443.4 & $< 22.3$ & near \m31 centre\\
\noalign{\smallskip}
 2009-10b& 00:42:20.77 & 55113.99 & mrg4 (HRC-I) & 400.2 & $< 8.4$ & \\
        & +41:16:44.5 &      & mrg4 (EPIC) & 442.4 & $< 0.1$ & \\
\noalign{\smallskip}
 2009-11b& 00:42:39.61 & 55135.13 & mrg4 (HRC-I) & 379.0 & $< 19.3$ & off-axis\\
\end{tabular}
\end{center}
\end{table*}
%

\begin{table*}[t!]
\begin{center}
\addtocounter{table}{-1}
\caption[]{continued.}
\begin{tabular}{lrrrrrl}
\hline\noalign{\smallskip}
\hline\noalign{\smallskip}
\multicolumn{3}{l}{Optical nova candidate} & \multicolumn{4}{l}{X-ray measurements} \\
\noalign{\smallskip}\hline\noalign{\smallskip}
\multicolumn{1}{l}{Name} & \multicolumn{1}{c}{RA~~~(h:m:s)$^a$} & \multicolumn{1}{c}{MJD$^b$}
& \multicolumn{1}{c}{Observation$^d$} & \multicolumn{1}{c}{$\Delta t^e$} & \multicolumn{1}{c}{$L_{\rm 50}^f$}
& \multicolumn{1}{l}{Comment} \\
M31N & \multicolumn{1}{c}{Dec~(d:m:s)$^a$} & \multicolumn{1}{c}{(d)} & \multicolumn{1}{c}{ID} &\multicolumn{1}{c}{(d)} &\multicolumn{1}{c}{(10$^{36}$ erg s$^{-1}$)} & \\ 
\noalign{\smallskip}\hline\noalign{\smallskip}
 2011-02a& 00:42:02.33 & 55607.13 & 13178 (HRC-I) & 2.0 & $< 31.2$ & off-axis \\
        & +41:29:15.6 &      & 13179 (HRC-I) & 12.1 & $< 31.1$ & \\
        &      &      & 13180 (HRC-I) & 23.0 & $< 32.2$ & \\
\noalign{\smallskip}
 2011-02c& 00:42:44.25 & 55615.78 & 13179 (HRC-I) & 3.5 & $< 1.7$ & \\
        & +41:16:49.3 &      & 13180 (HRC-I) & 14.3 & $< 1.2$ & \\
\noalign{\smallskip}
\hline
\noalign{\smallskip}
\end{tabular}
\end{center}
Notes: \hspace{0.2cm} As for Table\,\ref{tab:novae_old_lum}.\\
\end{table*}
%

\begin{table*}[t!]
\begin{center}
\caption[]{Upper limits for \m31 CNe with outburst from about one year before the start of the 2011/12 monitoring until its end.}
\begin{tabular}{lrrrrrl}
\hline\noalign{\smallskip}
\hline\noalign{\smallskip}
\multicolumn{3}{l}{Optical nova candidate} & \multicolumn{4}{l}{X-ray measurements} \\
\noalign{\smallskip}\hline\noalign{\smallskip}
\multicolumn{1}{l}{Name} & \multicolumn{1}{c}{RA~~~(h:m:s)$^a$} & \multicolumn{1}{c}{MJD$^b$}
& \multicolumn{1}{c}{Observation$^d$} & \multicolumn{1}{c}{$\Delta t^e$} & \multicolumn{1}{c}{$L_{\rm 50}^f$}
& \multicolumn{1}{l}{Comment} \\
M31N & \multicolumn{1}{c}{Dec~(d:m:s)$^a$} & \multicolumn{1}{c}{(d)} & \multicolumn{1}{c}{ID} &\multicolumn{1}{c}{(d)} &\multicolumn{1}{c}{(10$^{36}$ erg s$^{-1}$)} & \\ 
\noalign{\smallskip}\hline\noalign{\smallskip}
 2010-10a& 00:42:45.84 & 55473.79 & mrg5 (HRC-I) & 403.3 & $< 26.7$ & off-axis\\
        & +41:24:22.2 &      & mrg5 (EPIC) & 449.3 & $< 0.2$ & \\
\noalign{\smallskip}
 2010-10f& 00:42:43.58 & 55481.19 & mrg5 (HRC-I) & 395.9 & $< 1.8$ & \\
        & +41:12:42.6 &      & mrg5 (EPIC) & 441.9 & $< 0.1$ & \\
\noalign{\smallskip}
 2010-10d& 00:42:36.91 & 55497.72 & mrg5 (HRC-I) & 379.4 & $< 1.4$ & \\
\end{tabular}
\end{center}
Notes: \hspace{0.2cm} As for Table\,\ref{tab:novae_old_lum}.\\
\end{table*}
%

\begin{table*}[t!]
\begin{center}
\addtocounter{table}{-1}
\caption[]{continued.}
\begin{tabular}{lrrrrrl}
\hline\noalign{\smallskip}
\hline\noalign{\smallskip}
\multicolumn{3}{l}{Optical nova candidate} & \multicolumn{4}{l}{X-ray measurements} \\
\noalign{\smallskip}\hline\noalign{\smallskip}
\multicolumn{1}{l}{Name} & \multicolumn{1}{c}{RA~~~(h:m:s)$^a$} & \multicolumn{1}{c}{MJD$^b$}
& \multicolumn{1}{c}{Observation$^d$} & \multicolumn{1}{c}{$\Delta t^e$} & \multicolumn{1}{c}{$L_{\rm 50}^f$}
& \multicolumn{1}{l}{Comment} \\
M31N & \multicolumn{1}{c}{Dec~(d:m:s)$^a$} & \multicolumn{1}{c}{(d)} & \multicolumn{1}{c}{ID} &\multicolumn{1}{c}{(d)} &\multicolumn{1}{c}{(10$^{36}$ erg s$^{-1}$)} & \\ 
\noalign{\smallskip}\hline\noalign{\smallskip}
  2011-11c& 00:41:58.24 & 55877.73 & 13228 (HRC-I) & 8.5 & $< 20.1$ & off-axis\\
        & +41:24:01.1 &      & 13229 (HRC-I) & 18.2 & $< 21.2$ & \\
        &      &      & 13230 (HRC-I) & 28.8 & $< 23.0$ & \\
        &      &      & 13231 (HRC-I) & 37.6 & $< 27.3$ & \\
        &      &      & mrg5 (EPIC) & 45.3 & $< 0.3$ & \\
\end{tabular}
\end{center}
Notes: \hspace{0.2cm} As for Table\,\ref{tab:novae_old_lum}.\\
\end{table*}
%

%
\begin{table}
\begin{center}
\caption[]{Non-nova SSSs in the monitoring.}
\begin{tabular}{rrl}
\hline\noalign{\smallskip}
\hline\noalign{\smallskip}
\multicolumn{1}{c}{Position (J2000.)} & ID$^a $ & Comment\\
\noalign{\smallskip}\hline\noalign{\smallskip}
  00:43:18.90 +41:20:17.9 & 1194 & strongly variable (\mzk)\\
  00:43:18.72 +41:18:07.0 & (*)875 & only found in 2010/11\\
  00:42:52.49 +41:15:40.7 & 1061 & bright SSS with period\\
  00:42:18.48 +41:26:46.2 & 857 & faint, nearby neighbour\\
  00:42:47.16 +41:14:13.5 & 1025 & faint, SSS in merge 2011/12\\
\noalign{\smallskip} \hline
\end{tabular}
\label{tab:sss}
\end{center}
Notes:\hspace{0.1cm} $^a $: Source number in the catalogue of \citet{2011A&A...534A..55S}, except for (*), which was only present in the work of \citet{2008A&A...480..599S}.\\
\end{table}

\clearpage

\begin{landscape}
\begin{table*}
\hspace{-8cm}
\begin{minipage}[b]{25.25cm}
\caption{Catalogue of X-ray detected optical novae in \m31.}
\label{tab:cat}
\begin{center}
\begin{tabular}{llllllrllllrrl}
\hline\noalign{\smallskip}
\hline\noalign{\smallskip}
\multicolumn{7}{l}{\normalsize{Optical measurements}} &
\multicolumn{4}{l}{\normalsize{X-ray measurements}} &
\multicolumn{2}{l}{\normalsize{Derived parameters}} &
\multicolumn{1}{l}{\normalsize{References}}\\
\noalign{\smallskip}\hline\noalign{\smallskip}
\multicolumn{1}{c}{Name} &
\multicolumn{1}{c}{Outburst$^a$} &
\multicolumn{1}{c}{Brightness$^b$} &
\multicolumn{1}{c}{$t_{\rm 2R}$ $^c$} &
\multicolumn{1}{c}{Old/$^d$} &
\multicolumn{1}{c}{Type$^e$} &
\multicolumn{1}{c}{$v_{\mbox{exp}}$ $^f$} &
\multicolumn{1}{c}{Turn-on} &
\multicolumn{1}{c}{Turn-off} &
\multicolumn{1}{c}{SSS$^g$/} &
\multicolumn{1}{c}{$kT_{\rm BB}^h$} &
\multicolumn{1}{c}{Ejected mass} &
\multicolumn{1}{c}{Burned mass} &
\multicolumn{1}{c}{o(ptical)$^i$} \\
\noalign{\smallskip}
\multicolumn{1}{c}{M31N} &
\multicolumn{1}{c}{(JD)} &
\multicolumn{1}{c}{(mag Filter)}& 
\multicolumn{1}{c}{(d)} &
\multicolumn{1}{c}{Young} &
\multicolumn{1}{c}{} &
\multicolumn{1}{c}{(km s$^{-1}$)} &
\multicolumn{1}{c}{(d)} & 
\multicolumn{1}{c}{(d)} &
\multicolumn{1}{c}{$kT$ date} & 
\multicolumn{1}{c}{(eV)} &
\multicolumn{1}{c}{($10^{-6}$ \msun)} & 
\multicolumn{1}{c}{($10^{-6}$ \msun)} &
\multicolumn{1}{c}{and x(-ray)$^j$}\\
\noalign{\smallskip}\hline\noalign{\smallskip}
\noalign{\medskip}	 
 1982-09b& 45224.51& 16.7(H$\alpha$)& & O& & & 2874& $3595\pm721$& &   & 10.92& $5.36\pm1.07$& o1;x1\\
\noalign{\medskip}     
 1990-09a& 48151.44& 15.7(H$\alpha$)& & Y& & & $156\pm156$& $501\pm188$& ROSAT&  & $4.32^{+25.77}_{-4.32}$& $1.55\pm0.58$& o2,3;x2,1\\
\noalign{\medskip}     
 1990-09g& 48161.00& 18.4(H$\alpha$)& & O& & & $152\pm152$& $842\pm538$& &   & $4.29^{+25.42}_{-4.29}$& $2.25\pm1.44$& o3;x1\\
\noalign{\medskip}     
 1990-12a& 48235.00& 16.0(H$\alpha$)& & O& & & $417\pm188$& $955\pm188$& &   & $5.91^{+20.32}_{-5.06}$& $2.46\pm0.48$& o3;x1,3\\
\end{tabular}
\end{center}
\renewcommand{\arraystretch}{1}
\end{minipage}
\end{table*}
\end{landscape}

\begin{landscape}
\begin{table*}
\addtocounter{table}{-1}
\hspace{-8cm}
\begin{minipage}[b]{25.25cm}
\caption{continued.}
\begin{center}
\begin{tabular}{llllllrllllrrl}
\hline\noalign{\smallskip}
\hline\noalign{\smallskip}
\multicolumn{7}{l}{\normalsize{Optical measurements}} &
\multicolumn{4}{l}{\normalsize{X-ray measurements}} &
\multicolumn{2}{l}{\normalsize{Derived parameters}} &
\multicolumn{1}{l}{\normalsize{References}}\\
\noalign{\smallskip}\hline\noalign{\smallskip}
\multicolumn{1}{c}{Name} &
\multicolumn{1}{c}{Outburst$^a$} &
\multicolumn{1}{c}{Brightness$^b$} &
\multicolumn{1}{c}{$t_{\rm 2R}$ $^c$} &
\multicolumn{1}{c}{Old/$^d$} &
\multicolumn{1}{c}{Type$^e$} &
\multicolumn{1}{c}{$v_{\mbox{exp}}$ $^f$} &
\multicolumn{1}{c}{Turn-on} &
\multicolumn{1}{c}{Turn-off} &
\multicolumn{1}{c}{SSS?$^g$/} &
\multicolumn{1}{c}{$kT_{\rm BB}^h$} &
\multicolumn{1}{c}{Ejected mass} &
\multicolumn{1}{c}{Burned mass} &
\multicolumn{1}{c}{o(ptical)$^i$} \\
\noalign{\smallskip}
\multicolumn{1}{c}{M31N} &
\multicolumn{1}{c}{(MJD)} &
\multicolumn{1}{c}{(mag Filter)}& 
\multicolumn{1}{c}{(d)} &
\multicolumn{1}{c}{Young} &
\multicolumn{1}{c}{} &
\multicolumn{1}{c}{(km s$^{-1}$)} &
\multicolumn{1}{c}{(d)} & 
\multicolumn{1}{c}{(d)} &
\multicolumn{1}{c}{$kT$ date} & 
\multicolumn{1}{c}{(eV)} &
\multicolumn{1}{c}{($10^{-6}$ \msun)} & 
\multicolumn{1}{c}{($10^{-6}$ \msun)} &
\multicolumn{1}{c}{and x(-ray)$^j$}\\
\noalign{\smallskip}\hline\noalign{\smallskip}
 2001-10b& 52190.98& 15.6(i')& 15.8& Y& & & $214\pm214$& & &  & $4.78^{+30.51}_{-4.78}$& & o9,10;x6\\
\noalign{\medskip}     
 2001-10f& 52195.76& 16.6(B)& 12.0& Y& & & $51\pm34$& $550\pm460$& 86& $51.0^{+4.0}_{-4.0}$  & $3.03^{+9.4}_{-2.8}$& $1.66\pm1.39$& o15,16;x1,5\\
\noalign{\medskip}     
 2001-11a& 52225.71& 17.1(B)& 8.0& Y& & & 53& $>100$& 53& $28.0^{+3.0}_{-3.0}$  & 3.07& $>0.44$& o17,16;x8,9\\
\noalign{\medskip}     
 2002-01b& 52281.81& 16.8(R)& 8.0& O& He/N& 1715 & $77\pm69$& $534\pm390$& &   & $11.19^{+29.04}_{-11.07}$& $1.63\pm1.19$& o18,19;x1\\
\end{tabular}
\end{center}
\renewcommand{\arraystretch}{1}
\end{minipage}
\end{table*}
\end{landscape}

\begin{landscape}
\begin{table*}
\addtocounter{table}{-1}
\hspace{-8cm}
\begin{minipage}[b]{25.25cm}
\caption{continued.}
\begin{center}
\begin{tabular}{llllllrllllrrl}
\hline\noalign{\smallskip}
\hline\noalign{\smallskip}
\multicolumn{7}{l}{\normalsize{Optical measurements}} &
\multicolumn{4}{l}{\normalsize{X-ray measurements}} &
\multicolumn{2}{l}{\normalsize{Derived parameters}} &
\multicolumn{1}{l}{\normalsize{References}}\\
\noalign{\smallskip}\hline\noalign{\smallskip}
\multicolumn{1}{c}{Name} &
\multicolumn{1}{c}{Outburst$^a$} &
\multicolumn{1}{c}{Brightness$^b$} &
\multicolumn{1}{c}{$t_{\rm 2R}$ $^c$} &
\multicolumn{1}{c}{Old/$^d$} &
\multicolumn{1}{c}{Type$^e$} &
\multicolumn{1}{c}{$v_{\mbox{exp}}$ $^f$} &
\multicolumn{1}{c}{Turn-on} &
\multicolumn{1}{c}{Turn-off} &
\multicolumn{1}{c}{SSS$^g$/} &
\multicolumn{1}{c}{$kT_{\rm BB}^h$} &
\multicolumn{1}{c}{Ejected mass} &
\multicolumn{1}{c}{Burned mass} &
\multicolumn{1}{c}{o(ptical)$^i$} \\
\noalign{\smallskip}
\multicolumn{1}{c}{M31N} &
\multicolumn{1}{c}{(MJD)} &
\multicolumn{1}{c}{(mag Filter)}& 
\multicolumn{1}{c}{(d)} &
\multicolumn{1}{c}{Young} &
\multicolumn{1}{c}{} &
\multicolumn{1}{c}{(km s$^{-1}$)} &
\multicolumn{1}{c}{(d)} & 
\multicolumn{1}{c}{(d)} &
\multicolumn{1}{c}{$kT$ date} & 
\multicolumn{1}{c}{(eV)} &
\multicolumn{1}{c}{($10^{-6}$ \msun)} & 
\multicolumn{1}{c}{($10^{-6}$ \msun)} &
\multicolumn{1}{c}{and x(-ray)$^j$}\\
\noalign{\smallskip}\hline\noalign{\smallskip}
 2006-09c& 53996.14& 17.0(R)& 23.0& Y& Fe II& 960  & $275\pm138$& $327\pm164$& 477& $74.0^{+14.0}_{-15.0}$ & $44.73^{+56.15}_{-33.63}$& $1.13\pm0.57$& o32,33;x11\\
\noalign{\medskip}     
 2006-11a& 54063.67& 16.0(R)& 29.0& Y& Fe II& 560 & $147\pm41$& $208\pm21$& 188& $65.0^{+25.0}_{-25.0}$ & $4.35^{+2.76}_{-2.09}$& $0.79\pm0.08$& o34,4;x13,14\\
\noalign{\medskip}     
 2007-02b& 54134.80& 16.7(R)& 34.0& Y& Fe II& 955  & $175\pm164$& $>1258$& 722& $32.0^{+7.0}_{-5.0}$ & $17.92^{+49.34}_{-17.85}$& $>2.95$& o35,36;x11,10\\
\end{tabular}
\end{center}
\renewcommand{\arraystretch}{1}
\end{minipage}
\end{table*}
\end{landscape}

\begin{landscape}
\begin{table*}
\addtocounter{table}{-1}
\hspace{-8cm}
\begin{minipage}[b]{25.25cm}
\caption{continued.}
\begin{center}
\begin{tabular}{llllllrllllrrl}
\hline\noalign{\smallskip}
\hline\noalign{\smallskip}
\multicolumn{7}{l}{\normalsize{Optical measurements}} &
\multicolumn{4}{l}{\normalsize{X-ray measurements}} &
\multicolumn{2}{l}{\normalsize{Derived parameters}} &
\multicolumn{1}{l}{\normalsize{References}}\\
\noalign{\smallskip}\hline\noalign{\smallskip}
\multicolumn{1}{c}{Name} &
\multicolumn{1}{c}{Outburst$^a$} &
\multicolumn{1}{c}{Brightness$^b$} &
\multicolumn{1}{c}{$t_{\rm 2R}$ $^c$} &
\multicolumn{1}{c}{Old/$^d$} &
\multicolumn{1}{c}{Type$^e$} &
\multicolumn{1}{c}{$v_{\mbox{exp}}$ $^f$} &
\multicolumn{1}{c}{Turn-on} &
\multicolumn{1}{c}{Turn-off} &
\multicolumn{1}{c}{SSS$^g$/} &
\multicolumn{1}{c}{$kT_{\rm BB}^h$} &
\multicolumn{1}{c}{Ejected mass} &
\multicolumn{1}{c}{Burned mass} &
\multicolumn{1}{c}{o(ptical)$^i$} \\
\noalign{\smallskip}
\multicolumn{1}{c}{M31N} &
\multicolumn{1}{c}{(MJD)} &
\multicolumn{1}{c}{(mag Filter)}& 
\multicolumn{1}{c}{(d)} &
\multicolumn{1}{c}{Young} &
\multicolumn{1}{c}{} &
\multicolumn{1}{c}{(km s$^{-1}$)} &
\multicolumn{1}{c}{(d)} & 
\multicolumn{1}{c}{(d)} &
\multicolumn{1}{c}{$kT$ date} & 
\multicolumn{1}{c}{(eV)} &
\multicolumn{1}{c}{($10^{-6}$ \msun)} & 
\multicolumn{1}{c}{($10^{-6}$ \msun)} &
\multicolumn{1}{c}{and x(-ray)$^j$}\\
\noalign{\smallskip}\hline\noalign{\smallskip}
 2010-12b& 55540.12& 15.7(R)& 3.0& Y& & & $21\pm5$& $219\pm164$& 26& $39.0^{+21.0}_{-30.0}$ & $2.28^{+2.72}_{-1.36}$& $0.83\pm0.62$& o63,64;x10\\
\noalign{\medskip}     
 2011-01a& 55568.39& 14.9(R)& 10.0& O& Fe II& 650 & $168\pm141$& $>355$& &  & $7.65^{+18.24}_{-7.45}$& $>1.2$& o65,66;x10\\
\noalign{\medskip}     
 2011-01b& 55577.52& 16.4(R)& 3.0& O& & & $25\pm7$& $248\pm51$& 197& $40.0^{+14.0}_{-20.0}$ & $2.41^{+3.22}_{-1.54}$& $0.91\pm0.19$& o68;x21,10\\
\noalign{\medskip}     
 2011-02b& 55615.78& 17.7(R)& & O& &  & & & &  & & & o68,34;x10\\
\end{tabular}
\end{center}
Notes:\hspace{0.2cm} $^a$: Modified Julian day of optical nova outburst; $^b $: maximum observed magnitude, ``W" indicates unfiltered magnitude; $^c$: time in days the nova R magnitude needs to drop 2 mag below peak magnitude \citep[see][]{1964gano.book.....P}; $^d$: positional association with the old (bulge) and young (disk) stellar populations of \m31 (see Sect.\,\ref{sec:discuss_pop}) ; $^e$: spectral type of optical nova according to the classification scheme of \citet{1992AJ....104..725W}; $^f$: outflow velocity of the ejected envelope as measured from optical spectra; $^g$: indicates if the source was classified as an SSS using \xmm hardness ratios (HR), \chandra HRC-I/ACIS-I hardness ratios (HR1), \chandra HRC-I hardness ratios (HR2), a ROSAT observation (ROSAT, only M31N~1990-09a), or using X-ray spectra, in which case we give the time in days after outburst for which $kT$ was determined (if multiple observations were used, this value is the mean of the associated days, if necessary weighted by flux); $^h$: maximum black body temperature as derived from spectral fits; $^i$: optical references: o1: \citet{1987ApJ...318..520C}, o2: \citet{2002A&A...389..439N}, o3: \citet{2001ApJ...563..749S}, o4: \citet{2011ApJ...734...12S}, o5: \citet{2004A&A...421..509A}, o6: \citet{1999AAS...195.3608R}, o7: \citet{2008A&A...477...67H}, o8: \citet{1998AstL...24..641S}, o9: \citet{2004MNRAS.353..571D}, o10: \citet{2006MNRAS.369..257D}, o11: \citet{1999IAUC.7272....2F}, o12: CBAT \m31 nova webpage (http://www.cfa.harvard.edu/iau/CBAT\_M31.html), o13: MPE \m31 nova catalogue (http://www.mpe.mpg.de/$\sim$m31novae/opt/m31/index.php), o14: \citet{2001IAUC.7738....3F}, o15: \citet{2006IBVS.5720....1S}, o16: \citet{2008AstL...34..563A}, o17: \citet{2006IBVS.5737....1S}, o18: \citet{2002IAUC.7794....1F}, o19: \citet{2002IAUC.7825....3F}, o20: \citet{2012A&A...537A..43L}, o21: \citet{2003IAUC.8226....2F}, o22: \citet{2003IAUC.8231....4D}, o23: \citet{2003IAUC.8248....2H}, o24: \citet{2007A&A...465..375P}, o25: D. Bishops extragalactic nova webpage (http://www.rochesterastronomy.org/novae.html), o26: \citet{2005ATel..421....1D}, o27: \citet{2006ATel..850....1P}, o28: \citet{2006ATel..805....1P}, o29: \citet{2006ATel..808....1B}, o30: \citet{2006ATel..821....1L}, o31: \citet{2006ATel..829....1R}, o32: \citet{2006ATel..887....1Q}, o33: \citet{2006ATel..923....1S}, o34: K. Hornoch (priv. comm.), o35: Burwitz et al. (2013, in prep.), o36: \citet{2007ATel.1009....1P}, o37: \citet{2007ApJ...671L.121S}, o38: \citet{2007ATel.1238....1B}, o39: \citet{2007ATel.1242....1R}, o40: \citet{2007ATel.1257....1P}, o41: \citet{2009ApJ...705.1056B}, o42: \citet{2007ATel.1332....1S}, o43: \citet{2007ATel.1336....1H}, o44: \citet{2007ATel.1341....1S}, o45: \citet{2008ATel.1602....1H}, o46: \citet{2008ATel.1563....1O}, o47: \citet{2008ATel.1580....1H}, o48: \citet{2009ATel.2147....1P}, o49: \citet{2009ATel.2165....1H}, o50: \citet{2009ATel.2171....1D}, o51: \citet{2010ATel.2435....1P}, o52: \citet{2010CBET.2187....3H}, o53: \citet{2010CBET.2305....1H}, o54: \citet{2010CBET.2319....1H}, o55: \citet{2010CBET.2411....1H}, o56: \citet{2010CBET.2472....1N}, o57: \citet{2012ApJ...752..133C}, o58: \citet{2010ATel.2909....1S}, o59: \citet{2010CBET.2573....1H}, o60: \citet{2010ATel.3001....1P}, o61: \citet{2010ATel.3006....1S}, o62: \citet{2013A&A...549A.120H}, o63: \citet{2010ATel.3076....1P}, o64: \citet{2010CBET.2582....4S}, o65: \citet{2011CBET.2631....1N}, o66: \citet{2011CBET.2631....6H}, o67: \citet{2011CBET.2631....7A}, o68: CBAT "Transient Objects Confirmation Page" (http://www.cbat.eps.harvard.edu/unconf/tocp.html), o69: \citet{2011ATel.3693....1O}, o70: \citet{2011ATel.3699....1S}, o71: \citet{2011ATel.3725....1B}, o72: \citet{2011ATel.3778....1S}, o73: \citet{2012ATel.4096....1H}; $^j$: X-ray references: x1: \citet{2005A&A...442..879P}, x2: \citet{2002A&A...389..439N}, x3: \citet{2006A&A...454..773P}, x4: \citet{2010AN....331..212S}, x5: \citet{2007A&A...465..375P}, x6: \citet{2008ATel.1672....1N}, x7: \citet{2010A&A...523A..89H}, x8: \citet{2005ApJ...634..314T}, x9: \citet{2006ApJ...643..356W}, x10: this work, x11: \citet{2011A&A...533A..52H}, x12: \citet{2008ATel.1390....1O}, x13: \citet{2007ATel.1116....1P}, x14: \citet{2008A&A...489..707V}, x15: \citet{2009A&A...500..769H}, x16: \citet{2009A&A...498L..13H}, x17: \citet{2011A&A...531A..22P}, x18: \citet{2012A&A...544A..44H}, x19: \citet{2010ATel.3038....1P}, x20: \citet{2013A&A...549A.120H}, x21: \citet{2011ATel.3441....1H}, x22: \citet{2012ATel.4511....1H}.\\
\renewcommand{\arraystretch}{1}
\end{minipage}
\end{table*}
\end{landscape}

\end{document}